\documentclass{aa}  

\usepackage{graphicx}
\usepackage{txfonts}
\usepackage{lineno}
\usepackage{epstopdf}
\usepackage{amstext}

\begin{document}

   \title{The $\gamma$-ray sky seen at X-ray energies I.}

   \subtitle{Searching for the connection between X-rays and $\gamma$-rays in \,\emph{Fermi} BL Lac objects}

   \author{E. J. Marchesini
          \inst{1,2,3,4,5}
          \and
          A. Paggi \inst{1}
        \and
        F. Massaro \inst{1}
        \and
        N. Masetti \inst{5,6}
        \and
        R. D'Abrusco \inst{7}
        \and
        I. Andruchow \inst{2,4}
        \and
        R. de Menezes \inst{1,8}
              }

   \institute{Dipartimento di Fisica, Universit\`a degli Studi di Torino, via Pietro Giuria 1, I-10125 Turin, Italy
\and Facultad de Ciencias Astron\'omicas y Geof\'isicas, Universidad Nacional de La Plata, Paseo del Bosque, B1900FWA, La Plata, Argentina
\and INFN -- Istituto Nazionale di Fisica Nucleare, Sezione di Torino, via Pietro Giuria 1, I-10125 Turin, Italy
\and Instituto de Astrof\'isica de La Plata, CONICET--UNLP, CCT La Plata, Paseo del Bosque, B1900FWA, La Plata, Argentina
\and INAF -- Osservatorio di Astrofisica e Scienza dello Spazio, via Gobetti 93/3, I-40129, Bologna, Italy
\and Departamento de Ciencias Fisicas, Universidad Andres Bello, Fernandez Concha 700, Las Condes, Santiago, Chile
\and Center for Astrophysics | Harvard \& Smithsonian , 60 Garden St, Cambridge (MA) 02138, USA
\and Universidade de S\~ao Paulo, Instituto de Astronomia, Geof\'isica e Ci\^encias Atmosf\'ericas, Departamento de Astronomia, S\~ao Paulo, SP 05508-090, Brazil
             }

   \date{}


  \abstract
{BL Lac objects are an extreme type of active galactic nuclei (AGNs) that belong to the largest population of $\gamma$-ray sources: blazars. This class of AGNs shows a double-bumped spectral energy distribution that is commonly described in terms of a synchrotron self-Compton (SSC) emission process, whereas the low-energy component that dominates their emission between the infrared and the X-ray band is tightly connected to the high-energy component that peaks in the $\gamma$-rays. Two strong connections that link radio and mid-infrared emission of blazars to the emission in the $\gamma$-ray band are well established. They constitute the basis for associating $\gamma$-ray sources with their low-energy counterparts.}
   {We searched for a possible link between X-ray and $\gamma$-ray emissions for the subclass of BL Lacs using all archival Swift/XRT observations combined with Fermi data for a selected sample of 351 sources.}
{   Analyzing $\sim$2400 ks of Swift/XRT observations that were carried 
out until December 2018, we discovered that above the $\gamma$-ray flux threshold $F_{\gamma}\approx3\times10^{-12}\,\rm{erg}\,\rm{cm}^{-2}\,\rm{s}^{-1}$, 96\% of all \emph{Fermi} BL Lacs have an X-ray counterpart that is detected with signal-to-noise ratio higher than 3.}
{   We did not find any correlation or clear trend between X-ray and $\gamma$-ray fluxes and/or spectral shapes, but we discovered a correlation between the X-ray flux and the mid-infrared color. Finally, we discuss on a possible interpretation of our results in the SSC framework.}
   {}

   \maketitle
%

\section{Introduction} 
\label{sec:intro}
Blazars are a peculiar class of active galactic nuclei (AGNs) that is characterized by emission arising from a relativistic jet oriented at small angles \citep[e.g., less than a few degrees,][]{Lister13} with respect to the line of sight. This jet emission overwhelms most of the radiation of their host galaxy \citep{BlandfordRees78}.

Blazar emission is detected at all frequencies. It extends from radio \citep[see, e.g.,][]{Jorstad01,Ciaramella04,Ghirlanda10,Lister19} and even low radio frequencies \citep[see, e.g.,][for recent observational campaigns]{Nori14,Giroletti16}, infrared \citep[IR, see, e.g.,][]{Impey88,Stevens94,Massaro11,DAbrusco12} and optical \citep[see, e.g.,][for recent observational campaigns]{Carini92,Marchesini16,PenaHerazo17,Marchesini19} to X-rays \citep[see, e.g.,][]{Singh85,Giommi90,Sambruna96,Pian98,Paggi13,Landi15}. The emission is also detected in the $\gamma$-ray band \citep[see, e.g.,][]{Aharonian05,Albert07,Giannios09,Tavecchio11,Wehrle98,Ackermann15} and shows strong variability. It also has flaring states in which spectral shape and/or luminosities change \citep[see, e.g.,][]{Hartman01,Romero02,Bottcher07,Pandey17,Kaur17,Bruni18}.

Blazars can be classified into two main categories: flat-spectrum radio quasars, and BL Lac objects. Empirically, the distinction between these two classes is based on emission features in their optical spectra \citep{Stickel91}. The first class shows strong and broad emission lines that are typical of normal quasars, while the spectra of the second class are almost featureless (i.e., emission lines with equivalent widths smaller than 5 \AA). We here adopt the nomenclature established by Roma-BZCat \citep{Massaro15}, where flat-spectrum radio quasars are labeled BZQs and BL Lac objects are BZBs.

Blazars are the dominant class of active galaxies in the $\gamma$-ray sky \citep{Hartman99,Mattox02,Massaro15b}. In particular, $\sim$56\% of all associated and classified $\gamma$-ray sources that have been detected by the \,\emph{Fermi} Large Area Telescope (\,\emph{Fermi}-LAT) four-year Point Source Catalog (3FGL) belong to this class \citep{Acero15}. In the past decade, follow-up spectroscopic campaigns \citep[see, e.g.,][]{Coso1,Opt5,Opt6,Opt7,Opt8} confirmed that most of the sources that were classified as "Blazar Candidates of Uncertain type" (BCUs), introduced in the 3FGL catalog, are indeed blazars of BL Lac type \citep{Massaro16b,AlvarezCrespo16}. The same situation occurs in the preliminary version of the latest release of the \,\emph{Fermi} catalog\footnote{https://arxiv.org/abs/1902.10045} \citep{Opt9}. Moreover, follow-up campaigns of unassociated or unidentified $\gamma$-ray sources (UGSs) have also shown that a large portion of them appear to be associated with blazars \citep{Opt1,Opt2,Opt3,Opt4}.

The spectral energy distribution (SED) of blazars shows two components: the low-energy component peaks between infrared and X-rays, and the high-energy component peaks between hard X-rays and the $\gamma$-ray band. The low-energy component is attributed to synchrotron emission arising from electrons that are accelerated in the blazar jets, and the high-energy component is due to the inverse Compton (IC) process \citep[see, e.g.,][for recent analyses]{Ghisellini85,Maraschi92,Massaro06,Tramacere07,Finke08,Paggi09b,Tramacere11}. For BZBs in particular, the two emission processes are strictly connected because seed photons for the IC emission are emitted by electrons through synchrotron radiation (i.e., the synchrotron self-Compton, or SSC, scenario).

Two different subclasses were originally defined for BZBs to distinguish them on the basis of the ratio of their radio- to X-ray flux \citep{Maselli10a}: low-energy peaked BL Lacs (i.e., LBLs) and high-energy peaked BL Lacs (i.e., HBLs). This ratio is defined as $\Phi_{\rm{XR}}=10^{-3}\,\frac{\rm{F}_{\rm{X}}}{\rm{S}_{1.4}\Delta\nu}$, where $\rm{F}_{\rm{SX}}$ is the X-ray flux from the ROSAT (short for\,\emph{R\"ontgensatellit}) survey \citep{Voges99} in the 0.1 to 2 keV band, $\rm{S}_{1.4}\Delta\nu$ is the radio flux density at 1.4 GHz multiplied by the band frequency width $\Delta\nu$. Thus, values of $\Phi_{\rm{XR}}$ greater than 1 point toward an HBL classification, while values lower than 1 indicate an LBL type of source. This corresponds to the limit $\frac{\rm{F}_{\rm{X}}}{\rm{S}_{1.4}}=1.0\times10^{-11}$.

A strong link between $\gamma$-ray and radio emission in blazars, known as the \emph{$\gamma$-\textup{to-radio connection}} \citep{Stecker93,Taylor07,Bloom08,Ghirlanda11}, was discovered soon after the launch of the Energetic Gamma Ray Experiment Telescope (EGRET) on board the Compton Gamma-ray satellite \citep{Fichtel93}. This link was later confirmed through \,\emph{Fermi} observations \citep{Mahony10,Ackermann11,Ghirlanda11,Cutini14,Lico14}.

Moreover, combining $\gamma$-ray and mid-infrared observations, the latter collected with the Wide-field Infrared Survey Explorer \citep[WISE;][]{Wright10}, a tight connection between their emissions in these two bands has also been discovered \citep{Massaro11,Massaro12b,DAbrusco13,Massaro16}. In particular, the $\gamma$-ray -- infrared connection is strongly related not only to the blazar power, but also to their spectral shapes in the two different energy ranges; this is expected given the theoretical interpretation of their SED. 

These two connections strongly stimulated follow-up campaigns to search for blazar-like counterparts of UGSs, for example, in the radio band \citep{Massaro13b,Nori14,Lico14,Giroletti16} and/or by applying statistical procedures to find them in the WISE catalogs \citep{DAbrusco13,Coso2,DAbrusco14,DAbrusco19}. On the other hand, X-ray follow-up observations have also been carried out to search for blazar-like counterparts of UGSs \citep[e.g.,][]{Mirabal09,Kataoka12,Stroh13,Paggi13,Masetti13b,Landi15,Paiano17a}, even if these counterparts not supported by any firmly established observational link or connection between the X-ray and $\gamma$-ray emission of blazars.

Based on the SSC scenario that underlies the interpretation of BL Lac SEDs, a link between X-ray and $\gamma$-ray emission
could be expected because emission of the low- and high-energy components is related to the same particle distribution. This is the main aim of the analysis presented here. We focus on BZBs, whose $\gamma$-ray emission is probably not significantly contaminated by inverse Compton radiation of seed photons that arises from regions outside the jet \citep{Sikora13}. 

We aim to determine the portion of $\gamma$-ray blazars that have an X-ray counterpart with respect to their $\gamma$-ray flux, and whether their $\gamma$-ray emission (i.e., flux and/or spectral shape) correlates with the X-ray emission, as occurs in the radio and in the mid-infrared bands. Our investigation will provide the necessary scientific background to support and justify ongoing \citep{Stroh13} and future X-ray follow-up campaigns of \,\emph{Fermi} UGSs. A detailed study of the general behavior of BZBs in the X-ray band is crucial to improve the selection of BZB candidates within a UGS sample. Our search for a connection between X-rays and  $\gamma$-rays is also supported by the fact that HBLs that are detected at TeV energies \citep[e.g.,][]{Piner14,Piner18} are generally among the brightest X-ray sources in the extragalactic sky and their X-ray emission is also linked to the mid-infrared emission \citep{Massaro13}.

To carry out our investigation, we analyzed observations from the X-ray Telescope (XRT) o board the Neil Gehrels \emph{Swift} Observatory, performed before mid-December 2018, of $\gamma$-ray BZBs observed by Fermi. We made this choice because the ROSAT survey is relatively shallow. Only $\sim$60\% percent of all known \,\emph{Fermi} blazars are listed in the ROSAT survey catalog \citep{Voges99}. Nevertheless, \emph{Swift} performs an X-ray follow-up campaign of \,\emph{Fermi} UGSs\footnote{https://www.\emph{Swift}.psu.edu/unassociated/} \citep{Falcone13,Stroh13,Falcone14}. An extensive database is therefore available.

This paper is organized as follows. In \S 2 we present the sample selection criteria, while in \S 3 we describe the Swift/XRT data reduction procedures. \S 4 is devoted to results of our analysis, and in \S 5 we describe a possible interpretation of our results within the SSC framework. Finally, \S 6 is dedicated to a brief summary and our main conclusions.

Unless stated otherwise and throughout the whole paper, we adopted cgs units and a flat cosmology with $H_0=\,72\,\rm{km}\,\rm{s}^{-1}\,\rm{Mpc}^{-1}$, and $\Omega_{\Lambda}=0.74$ \citep{Dunkley09}. Spectral indices $\alpha$ were defined so that the flux density $\rm{S}_{\nu}\propto\nu^{-\alpha}$, considering $\alpha<0.5$ as \emph{\textup{flat spectra}}. The AllWISE magnitudes in the [3.4]$\mu$m, [4.6]$\mu$m, and [12]$\mu$m nominal filters are in the Vega system, and are not corrected for Galactic extinction because this correction is negligible for Galactic latitudes $|b|>10^{\circ}$ \citep[see, e.g.,][]{DAbrusco13}.

\section{Sample selection} \label{sec:sample}
To assess whether the X-ray and $\gamma$-ray emission in BZBs is connected, we started by selecting all known \,\emph{Fermi} BZBs listed in the ``clean'' sample of the Third Catalog of Active Galactic Nuclei detected by the\,\emph{Fermi}-LAT \cite[3LAC]{Ackermann15}, and considered only those that belong to the fifth release of the Roma-BZCat \citep{Massaro15}. At this selection step our sample contained 580 of the original 1151 sources. All BZBs in BZCat have a counterpart in at least one of the main radio surveys \citep{White97,Condon98,Mauch03}, and all selected BZBs are uniquely associated with $\gamma$-ray sources in the 3FGL catalog.

We only included in our sample \emph{Swift}/XRT observations that were performed in photon-counting (PC) mode that lay within a circular region of 6 arcmin angular separation around the BZB $\gamma$-ray positions (431 sources). Our choice of 6 arcmin corresponds to the average semimajor axis of the positional uncertainty ellipse of $\gamma$-ray sources listed in the 3FGL \citep{Acero15}. We chose only sources with total exposure times of between 1 and 20 ks because those with cumulative exposure times longer than 20 ks are generally pointed as follow-up observations of flaring states and are not snapshot observations. A similar criterion was adopted by \citet{Mao16}.

The \emph{Swift} X-ray campaign of UGSs is performed with a nominal 5 ks exposure time \citep{Stroh13}, implying that the results achieved for our selected sample are suitable to carry out a future investigation of UGSs observed with Swift/XRT (Marchesini et al. 2019, in prep.). Most of the observed fields have a 5 ks exposure time; the average of our sample is 6.7 ks.

\section{Swift/XRT data reduction and analysis}

\subsection{Data processing} \label{sec:data}

We adopted the same data reduction procedure as described in \citet{Massaro08,Massaro08b,Paggi13,Massaro11b,Massaro12c}. Here we report the basic details and highlight differences and improvements with respect to our previous analyses. 

Raw SWIFT/XRT data were download and reduced to obtain clean event files with the standard procedures using the \textsc{xrtpipeline} task, which is part of the \emph{Swift} X-Ray Telescope Data Analysis Software \citep[\textsc{XRTDAS},][]{Capalbi05}, and the  High Energy Astrophysics Science Archive Research Center (HEASARC) calibration database (\textsc{CALDB}) version \textsc{x20180710}. In particular, using \textsc{xselect,} we excluded time intervals with count rates exceeding 40 counts per second, and time intervals where the CCD temperature exceeds -50$^{\circ}$C in regions located at the CCD edge \citep{DElia13}.

Clean event files were merged using the \textsc{xselect} task, while corresponding exposure maps were merged with \textsc{ximage} software. Figure 1 shows a merged image obtained for the XRT field associated with 5BZB\,J2005+7752 as an example of the final product obtained with our code.

\begin{figure*}[ht!]
\centering
\centering \includegraphics[scale=0.65]{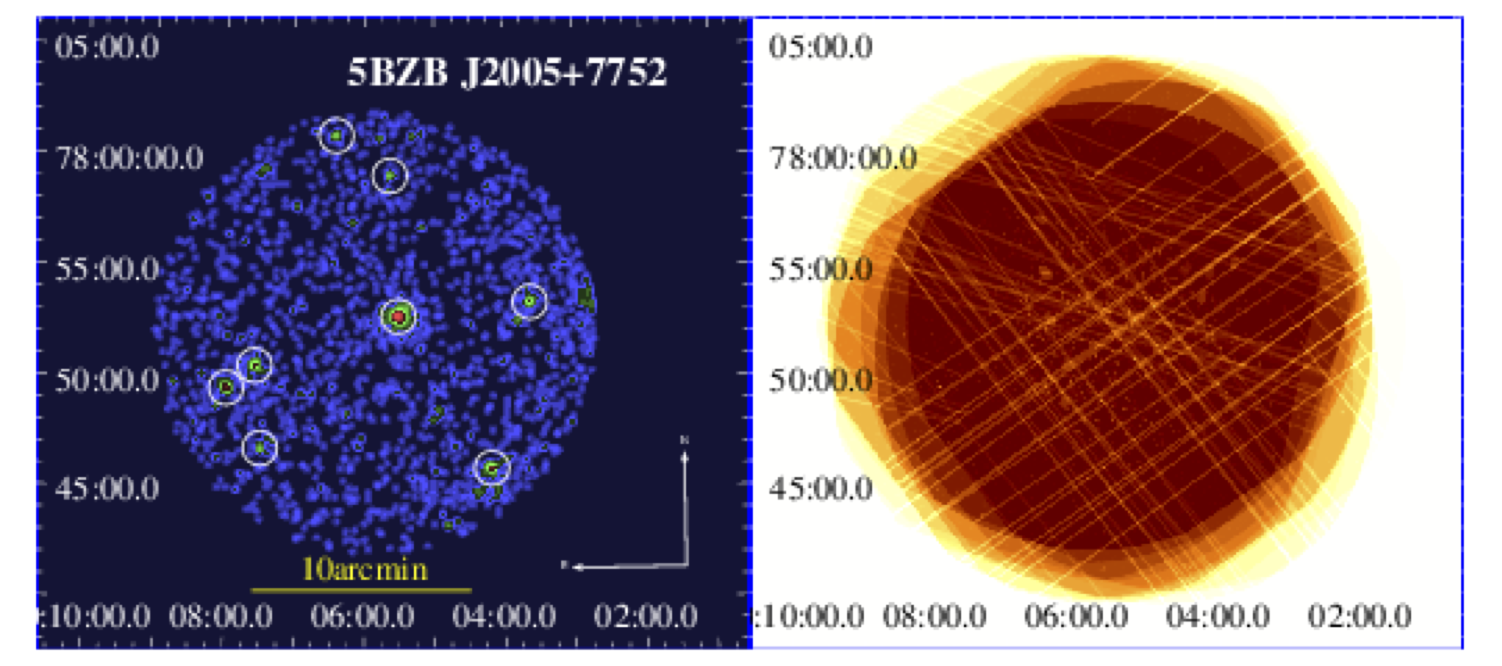}
\caption{X-ray image in the 0.5-10 keV energy range obtained by merging all XRT observations within 6 arcmin measured from the $\gamma$-ray position of 5BZB J2005+7752 (left panel) and its corresponding merged exposure map (right panel). The bright source in the center is the X-ray counterpart of the BZB, marked with the white circle as all other background and foreground X-ray sources. This X-ray image has been built by merging five observations for an exposure time of 15.9 ks. The image is also smoothed with a Gaussian kernel with a radius of 5 pixels.}
\end{figure*}

\subsection{Source detection and photometry}

A first detection run was performed over merged event files using the \textsc{DET} algorithm in \textsc{ximage}, to obtain pixel positions for every detection with signal-to-noise ratios (S/N) higher than 3. Because the exposure times chosen to carry out our investigation were relatively short, we focused on a photometric analysis.

We ran the \textsc{SOSTA} task available within the \textsc{XIMAGE} package on the pixel positions obtained from \textsc{DET}. In particular, \textsc{SOSTA} takes into account the local background for each source to claim a detection, which achieves a more precise photometry than the \textsc{DET} algorithm. This was carried out on merged event files for the full 0.5-10 keV band, and also for the soft (0.5-2 keV) and hard (2-10 keV) bands. Sources detected with \textsc{SOSTA}  all had an S/N between 3 and 25.

We compared the resulting X-ray sources with those listed in the \emph{Swift}-XRT point source (1SXPS) catalog \citep{Evans14}. Our procedure differs from that of 1SXPS in i) the choice of the S/N threshold applied to claim a detection, which is S/N$>$1.6, ii) the total number of Swift/XRT 
observations processed (1SXPS used those up to October 2012, while we 
reduced those up to December 2018), and iii) the choice of background regions (whose shape and size depend on the S/N of the source in 1SXPS). However, we found that our results agree with the 1SXPS catalog with differences of only a few percent (i.e., less than 5\%) mainly due to the reasons highlighted above.

Thus, we obtained positions, counts, and count rates for all sources. We flagged sources with full-band count rates higher than 0.5 photons per second, indicating the presence of pile-up. These are 13 sources, all of them with mild pile-up (i.e., with count rates lower than one photon per second). We then derived the hardness ratio ($\rm{HR}_{\rm{X}}$) for each source and full-band X-ray fluxes ($\rm{F}_{\rm{X}}$). The hardness ratio was computed as $(\rm{H}-\rm{S})/(\rm{H}+\rm{S})$, where $\rm{H}$ are counts in the \emph{\textup{hard}} band (2 to 10 keV) and $\rm{S}$ those in the \emph{\textup{soft}} band (0.5 to 2 keV). We verified that using counts or count rates to obtain $\rm{HR}_{\rm{X}}$ is equivalent because the exposure map is not energy dependent in the 0.5-10 keV band. Fluxes were obtained for each source using \textsc{PIMMS} \citep{Mukai93}, assuming a power law with a photon index of 2.0 and values of the Galactic column density values obtained from the LAB Survey of Galactic HI \citep{Kalberla05}. The choice of the photon index affects the estimate of the X-ray flux by a factor of less than a few percent \citep{Massaro11c,Massaro12d}. In Table 1 we report the X-ray data and information available from cross-matches 
that were performed with multifrequency archives (see Section 4.1 for details). In Col. 1 we report the BZCat name, in Cols. 2 to 4 the total counts and their error for the soft (0.5-2.0 keV), hard (2.0-10 keV), and total (0.5-10 keV) bands, in Cols. 5 to 10 the AllWISE magnitudes and their errors, and in Col. 11 the ratio of the X-ray to radio flux.

\section{Results}

\subsection{X-ray counterparts of $\gamma$-ray BL Lac objects}

In total, we found 1362 X-ray sources in the 351 BZB fields we reduced and analyzed. In Figure 2 we show the distribution of the total exposure times of our final sample, of all observed fields, considering those with at least one X-ray detection and those with no X-ray sources above an S/N of 3, separately. In 14 BZBs fields no XRT counterpart was found. All these sources were observed for a total exposure time shorter than 2.4 ks, with two exceptions: 5BZB\,J1458+3720 (5.3 ks) and 5BZB\,J0434-2342 (6.8 ks). The first does not show any detection in the XRT merged observation, and the second shows two X-ray sources at more than 6 arcmin from the BZB position. There is, however, a marginal detection with an S/N  of 2.8 coincident with the BZB position. This is also detected in the 1SXPS catalog with an S/N of 1.9. Both were discarded because no detection above the S/N threshold of 3 was reported. Thus the lack of X-ray counterparts for these 14 cases is mainly imputable to our choice of the S/N threshold combined with short exposure time.

\begin{figure}[ht!]
\centering \includegraphics[height=7.2cm,width=8.4cm,angle=0]{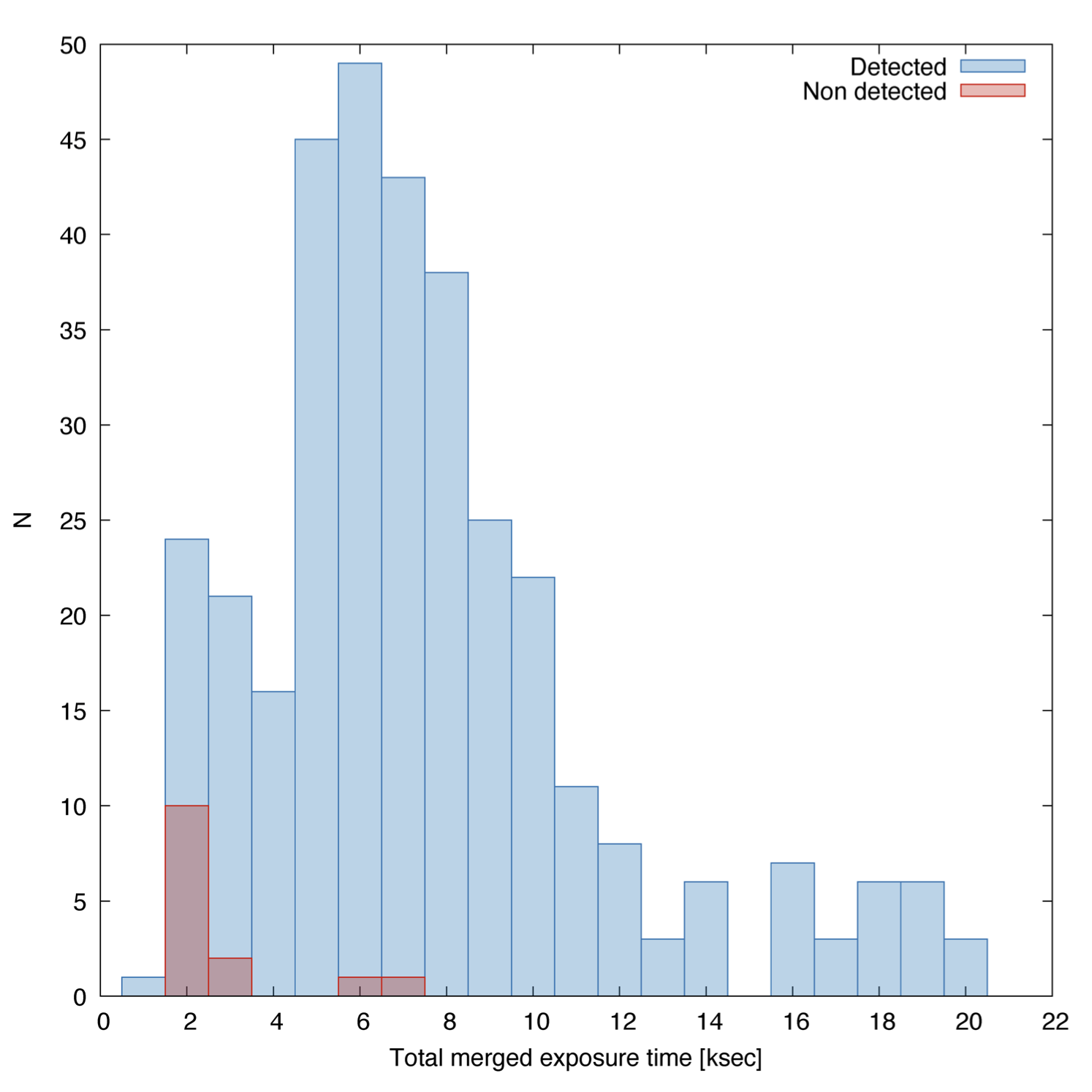}
\caption{Distribution of the total exposure time for all selected BZBs. The blue histogram indicates the X-ray merged event files with at least one X-ray source detected with an S/N greater than 3, while in red we show those without an X-ray detection.}
\end{figure}

The following cuts were also performed \emph{\textup{a posteriori}}:

\begin{enumerate}
    \item Spurious X-ray detections due to artifacts and bad pixels or bad columns were discarded.
    \item X-ray sources that were detected close ((i.e., closer than 8 arcsec) to a bright source were discarded as artifacts of the detection algorithm. The only exception was the source with the highest S/N.
    \item Sources that were clearly extended with respect to the XRT 90\% point spread function (PSF) \citep[23 arcsec,][]{Moretti04} were discarded.
    \item Sources not coincident with the BZB position (i.e., beyond 5.5 arcsec) and with negative counts on either soft and hard band (defined in Sec. 3.2) were discarded.
    \item 5BZB\,J1104+3812 (also known as Mrk 421), which presented severe pile-up contamination, was also discarded. Mrk 421 has been thoroughly studied in $\gamma$-rays and X-rays \citep[e.g.,][and references therein]{Brinkmann05,Isobe10,Banerjee19,Hervet19}.
    \item Two BZBs that lie in the same field within the positional uncertainty region of the same \,\emph{Fermi} source, 3FGL\,J0323.6-0109 (i.e., 5BZB\,J0323-0111 and 5BZB\,J0323-0108), were also discarded because of possible source confusion in $\gamma$-rays.
\end{enumerate}

In addition, we also cross-matched the X-ray positions with the AllWISE catalog with an uncertainty radius of 3.3 arcsec, following \citet{DAbrusco13}. All mid-infrared counterparts were correctly associated with the XRT detections, with two exceptions: 5BZB\,J0335-4459 (at 3.8 arcsec) and 5BZB\,J2131-2515 (at 3.5 arcsec)\footnote{5BZBJ\,1046-2535 was not included because of light contamination from a nearby star. 5BZB\,J2108-6637 is the only source for which no WISE counterpart was found.}. Because they are the closest WISE sources and unique matches, we kept these two BZBs in our sample.

Of the 351 Fermi BZBs observed by \emph{Swift}/XRT for 1 to 20 ks, 337 were detected and 14 were not. In addition, 3 BZBs were discarded because of pile-up or source confusion. Thus, we built a clean sample of 334 BZBs with X-ray and mid-infrared counterparts, and a clean sample of 675 background and foreground X-ray sources lying within an angular separation of 6 arcmin from the $\gamma$-ray position. The flow chart shown in Figure 3 summarizes all our selection steps.

\begin{figure}[ht!]
\centering
\centering \includegraphics[scale=0.90]{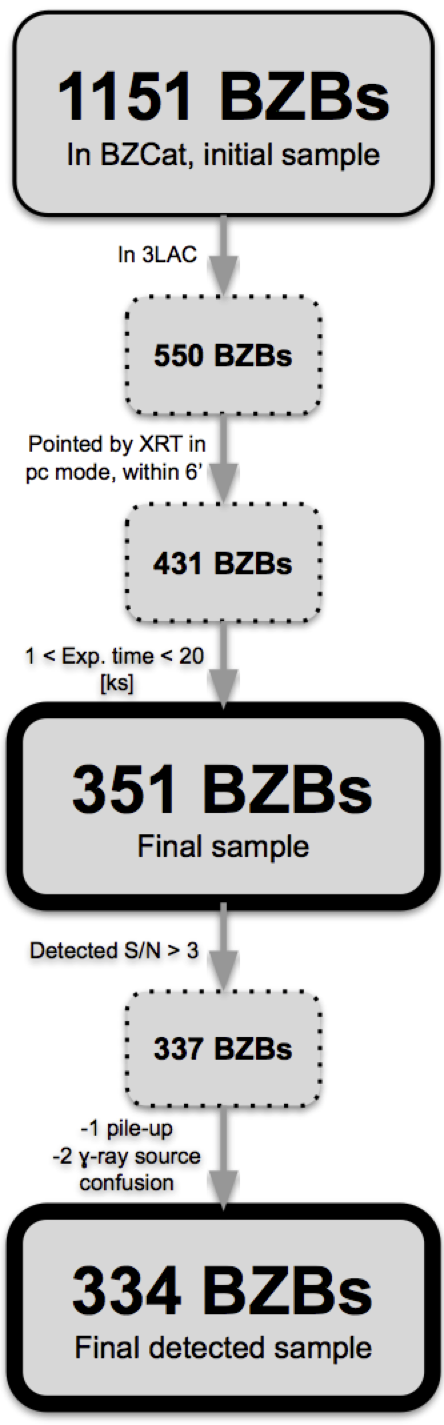}
\caption{Flow chart to highlight all steps we followed to build our final BZBs sample.}
\end{figure}

\subsection{X-ray properties of $\gamma$-ray BL Lac objects}

\label{sec:results}
We detected 337 BZBs using XRT data out of the original selected sample of 351 \,\emph{Fermi} BZBs. This means that 96\% of the BZB sources listed by \,\emph{Fermi} show an X-ray counterpart when they are observed for more than 1ks. This strongly supports the existence of a connection between X-rays and  $\gamma$-rays and certainly motivates X-ray follow-up observations at least to the level of the $\gamma$-ray flux of our current BZB sample to search for UGS counterparts \citep{Massaro13d}. 

In Figure 4 we show the distribution of the $\gamma$-ray energy flux in the 100 MeV to 100 GeV band for all \emph{Fermi} BZBs in our sample, divided into those for which we found at least one X-ray counterpart and those for which no XRT counterpart was detected. We plot the $\gamma$-ray flux thresholds above which we found at least one X-ray counterpart for 100\% ($\rm{F}_{\gamma}=1.7\times10^{-11}\,\rm{erg}\,\rm{cm}^{-2}\,\rm{s}$), 98\% ($\rm{F}_{\gamma}=7.0\times10^{-12}\,\rm{erg}\,\rm{cm}^{-2}\,\rm{s}$), and 96\% ($\rm{F}_{\gamma}=2.9\times10^{-12}\,\rm{erg}\,\rm{cm}^{-2}\,\rm{s}$) of the sample. We expect that these thresholds can be used to find new BZBs among the Fermi UGSs.

\begin{figure}[ht!]
\centering \includegraphics[height=7.2cm,width=8.4cm,angle=0]{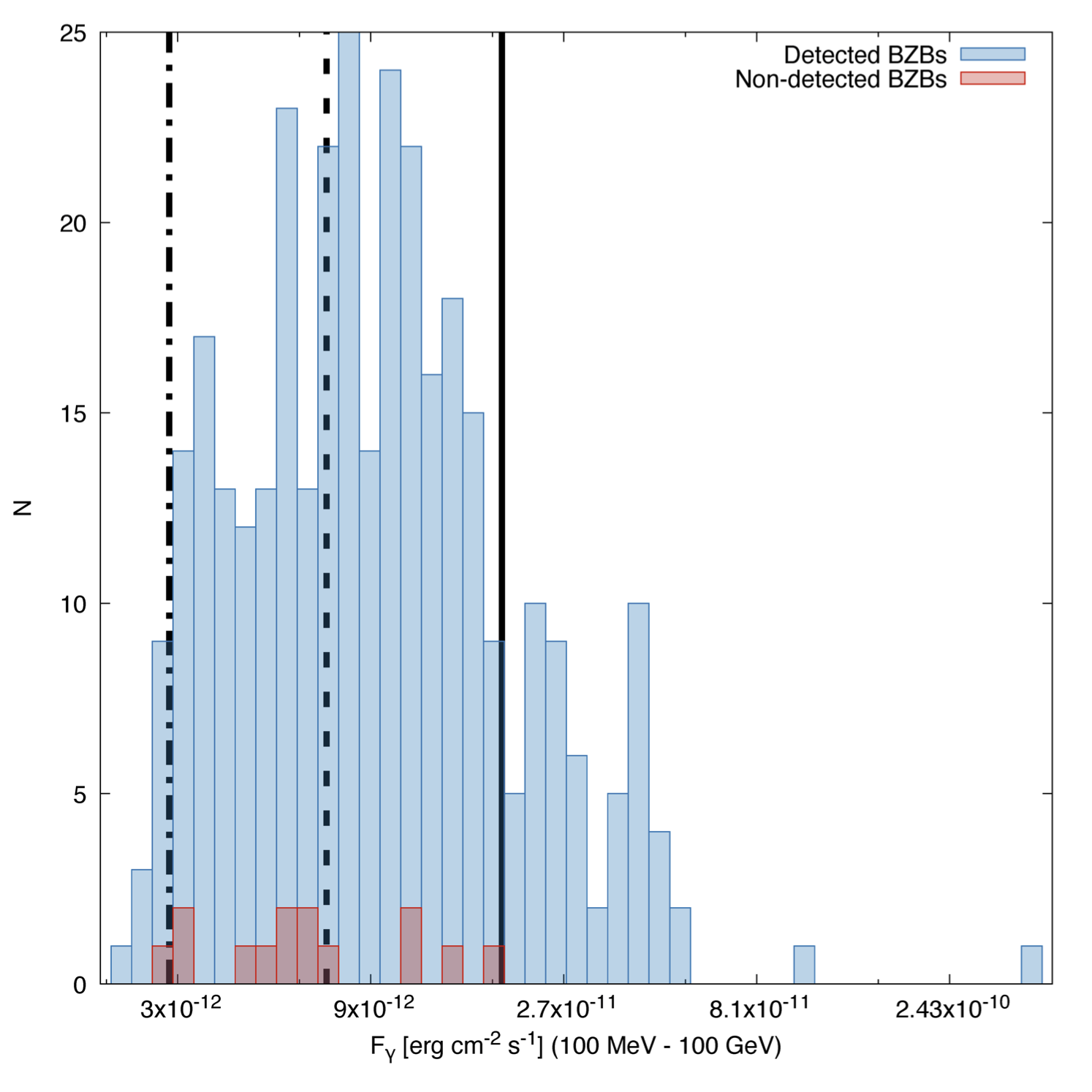}
\caption{Distribution of $F_{\gamma}$ in the 100 MeV to 100 GeV band for all 351 BZBs in our selected sample. As in Fig. 2, the blue histogram indicates the X-ray merged event file with at least one X-ray 
source detected with S/N greater than 3, and in red we show those without an X-ray detection. The solid black line corresponds to the threshold above which 100\% of the selected BZBs has an X-ray counterpart, while the dashed and dot-dashed black lines mark the 98\% and 96\% thresholds, respectively.}
\end{figure}

The nominal exposure time used for the UGS XRT follow-up campaign is 5 ks \citep{Stroh13}. Thus, we verified the fraction of BZBs that could be detected by rescaling the exposure time to this nominal value. We selected only sources for which the total (merged) exposure time was longer than 5 ks (222 sources), and then scaled their S/N assuming a Poisson distribution of the observed count rates. We confirm that 98\% of our selected sample would still be detected at an S/N ratio greater than 3 with shorter (i.e., 5ks) exposure time.

The HBL versus LBL classification was made following \citet{Maselli10a}, as mentioned in \S 1. The 1.4 GHz radio fluxes were taken from BZCat for each source, with the exception of 5BZB\,J1326-5256 and 5BZB\,J1604-4441, for which only the flux at 4.85 GHz was available. We assumed that the spectral shape of blazars is flat (radio spectral index equal to zero) at radio frequencies \citep[see, e.g.,][]{Healey07,Massaro13b,Massaro13c}, therefore we considered the 4.85 GHz flux to be equal to that at 1.4 GHz. Because the XRT band (0.5-10 keV) differs from that of ROSAT (0.1-2.4 keV), which is the band that was originally used for this classification, we compared the use of full-band fluxes adjusted to the ROSAT band assuming a power law. The HBL and LBL classification is the same using either of the X-ray flux estimates. Then we adopted full-band XRT fluxes. In Table 1 we show the full-band XRT-to-radio flux ratio $\rm{F}_{\rm{X}}/\rm{S}_{1.4}$ instead of $\Phi_{\rm{XR}}$, so that the limit between classes is $\rm{F}_{\rm{X}}/\rm{S}_{1.4}>10^{-11}$.

\begin{sidewaystable*}
\caption{First 10 rows for the main results of our analysis.} 
\centering
\begin{tabular}{lrrrrrrrrrr}
\hline
Source & Soft & Hard & Flux  &  W1 & $\sigma_{W1}$ & W2 & $\sigma_{W2}$ & W3 & $\sigma_{W3}$ & $\rm{F}_{\rm{X}}/\rm{S}_{1.4}$ \\
       & [photons] & [photons] & $[\times10^{-12}\rm{erg}\,\rm{cm}^{-2}\,\rm{s}^{-1}]$ &  [mag] & [mag] & [mag] & [mag] & [mag] & [mag] & $[\times10^{-12}]$\\
\hline
  5BZB\,J0001-0746 & $22.3\pm5.5$ & $11.4\pm3.9$      & $0.3\pm0.05$ & 12.68  & 0.01  & 11.76  & 0.01  & 9.17   & 0.04  & 1.4 \\
  5BZB\,J0004-1148 & $55.7\pm8.2$ & $13.9\pm4.3$      & $0.5\pm0.06$ & 14.10  & 0.02  & 13.12  & 0.04  & 10.4   & 0.1   & 1.0 \\
  5BZB\,J0008-2339 & $533.7\pm26.0$ & $122.2\pm13.0$  & $5.5\pm0.2$ & 13.75 & 0.02 & 13.29 & 0.03 & 11.3 & 0.2         & 150 \\
  5BZB\,J0009+0628 & $69.6\pm10.0$ & $23.4\pm6.1$     & $0.5\pm0.06$ & 12.97  & 0.01 & 12.04 & 0.02 & 9.36   & 0.04     & 2.1 \\
  5BZB\,J0009+5030 & $44.6\pm7.5$ & $9.1\pm3.4$       & $0.4\pm0.05$ & 12.97 & 0.02 & 12.44 & 0.02 & 10.37 & 0.08       & 29 \\
  5BZB\,J0014-5022 & $1051.0\pm37.0$ & $293.4\pm19.0$ & $7.6\pm0.2$ & 14.74 & 0.03 & 14.6 & 0.1 & 11.9 & (...)             & 590\\
  5BZB\,J0019+2021 & $15.9\pm4.7$ & $5.8\pm2.8$       & $0.2\pm0.05$ & 14.44  & 0.04  & 13.39 & 0.06 & 10.4   & 0.1     & 0.2 \\
  5BZB\,J0019-8152 & $91.3\pm11.0$ & $9.8\pm4.0$      & $0.6\pm0.06$ & 12.209 & 0.007 & 11.331 & 0.007 & 8.98  & 0.02   & 6.8 \\
  5BZB\,J0021-2550 & $35.9\pm6.9$ & $4.9\pm2.8$       & $0.2\pm0.03$ & 13.38 & 0.02 & 12.54 & 0.02 & 10.24 & 0.08       & 3.0 \\
  5BZB\,J0022+0608 & $67.3\pm8.9$ & $19.8\pm5.2$      & $0.4\pm0.05$ & 12.92 & 0.02 & 11.88 & 0.01 & 9.08  & 0.04       & 1.2 \\
\end{tabular}
\tablefoot{In Col. 1 we report the source name as listed in BZCat, in Cols. 2 and 3 the total counts and their error for the soft (0.5-2.0 keV) and hard (2.0-10 keV) bands, in Col. 4 the total band (0.5-10 keV) flux and its uncertainty, in $\rm{erg}\,\rm{cm}^{-2}\,\rm{s}^{-1}$ units, in columns 5 to 10 the AllWISE magnitudes and their uncertainties for each source, and in column 11 the X-ray to radio flux ratio. The AllWISE magnitudes W1, W2, and W3 correspond to wavelengths 3.4 $\mu$m, 4.6 $\mu$m, and 12 $\mu$m, respectively. Sources with pile-up are marked with a $ \text{dagger }$ next to their name. The complete table is available  in electronic form at the CDS via anonymous ftp to cdsarc.u-strasbg.fr (130.79.128.5)
or via http://cdsweb.u-strasbg.fr/cgi-bin/qcat?J/A+A/.}
\end{sidewaystable*}

In total, 175 sources out of the total 334 are HBLs, while the remaining 159 are classified as LBLs. As expected, errors in $\rm{HR}_{\rm{X}}$ and $\rm{F}_{\rm{X}}$ for HBL and LBL objects are smaller than for background or foreground objects because BZBs are generally one order of magnitude brighter (average flux $\rm{F}_{\rm{X}}=3.9\times10^{-12}\,\rm{erg}\,\rm{cm}^{-2}\,\rm{s}^{-1}$) than background or foreground objects (average flux $\rm{F}_{\rm{X}}=2.7\times10^{-13}\,\rm{erg}\,\rm{cm}^{-2}\,\rm{s}^{-1}$). On the other hand, HBLs and LBLs have softer X-ray spectra than background or foreground objects, which span all the possible $\rm{HR}_{\rm{X}}$ values. In Figure 5 we show $\rm{F}_{\rm{X}}$ versus $\rm{HR}_{\rm{X}}$ plotted for all detected sources (i.e., BZBs and background or foreground X-ray objects ). BZBs are separated into HBLs and LBLs with a color code indicating their $\Phi_{\rm{XR}}$ value, which extends from green (HBLs) to orange (LBLs), while sources with $\Phi_{\rm{XR}}\approx1$ are marked in yellow. The average value for the $\rm{HR}_{\rm{X}}$ is $\rm{HR}_{\rm{X}}=-0.63\pm0.09$ for HBLs and $\rm{HR}_{\rm{X}}=-0.46\pm0.16$ for LBLs. Sources with $\rm{F}_{\rm{X}}>2.0\times10^{-11}\,\rm{erg}\,\rm{cm}^{-2}\,\rm{s}^{-1}$ (indicated with the black solid line) suffer from pile-up. The HBL and LBL subsamples lie in different regions, as do background or foreground objects. Intermediate sources (i.e., with $\Phi_{\rm{XR}}\approx1$) lie in between both regions, as expected. This is also true in all figures shown from this point on.

To quantify the separation between each subsample in the  $\rm{F}_{\rm{X}}$--$\rm{HR}_{\rm{X}}$ space, we performed a kernel density estimation (KDE) analysis, in order to estimate density thresholds for each subsample, following \citet{DAbrusco09,Laurino11,Massaro11,Massaro13d,DAbrusco19}, and references therein. In Figure 6 we show $\rm{F}_{\rm{X}}$ versus $\rm{HR}_{\rm{X}}$ for background or foreground objects and for BZBs separated into HBLs and LBLs, with contours obtained from the KDE analysis for each subsample at 70\%, 80\%, and 90\% isodensity levels drawn on the basis of the probability evaluated with the KDE. Although the subsamples overlap, the overlap is smallest when we consider a 90\% density level. Moreover, HBLs lie in a very distinct region at all density levels.

From Figures 5 and 6, it follows that the LBL and HBL classification following \citet{Maselli10a} coincides with our results obtained through a KDE analysis, with intermediate-type objects populating the overlapping isodensity area between both classes.
In Figure 7 we also show the distribution of the angular separation between the X-ray and the $\gamma$-ray positions in the same field. X-ray detected BZBs on average appear to be closer to their $\gamma$-ray counterpart than background X-ray sources in the examined fields. This is confirmed through a Kolmogorov-Smirnov test, which yields a negligible chance coincidence probability, or p-chance $\rm{p}$. This implies a low contamination by background or foreground sources.

\begin{figure}[ht!]
\centering 
\includegraphics[height=6.2cm,width=8.4cm,angle=0]{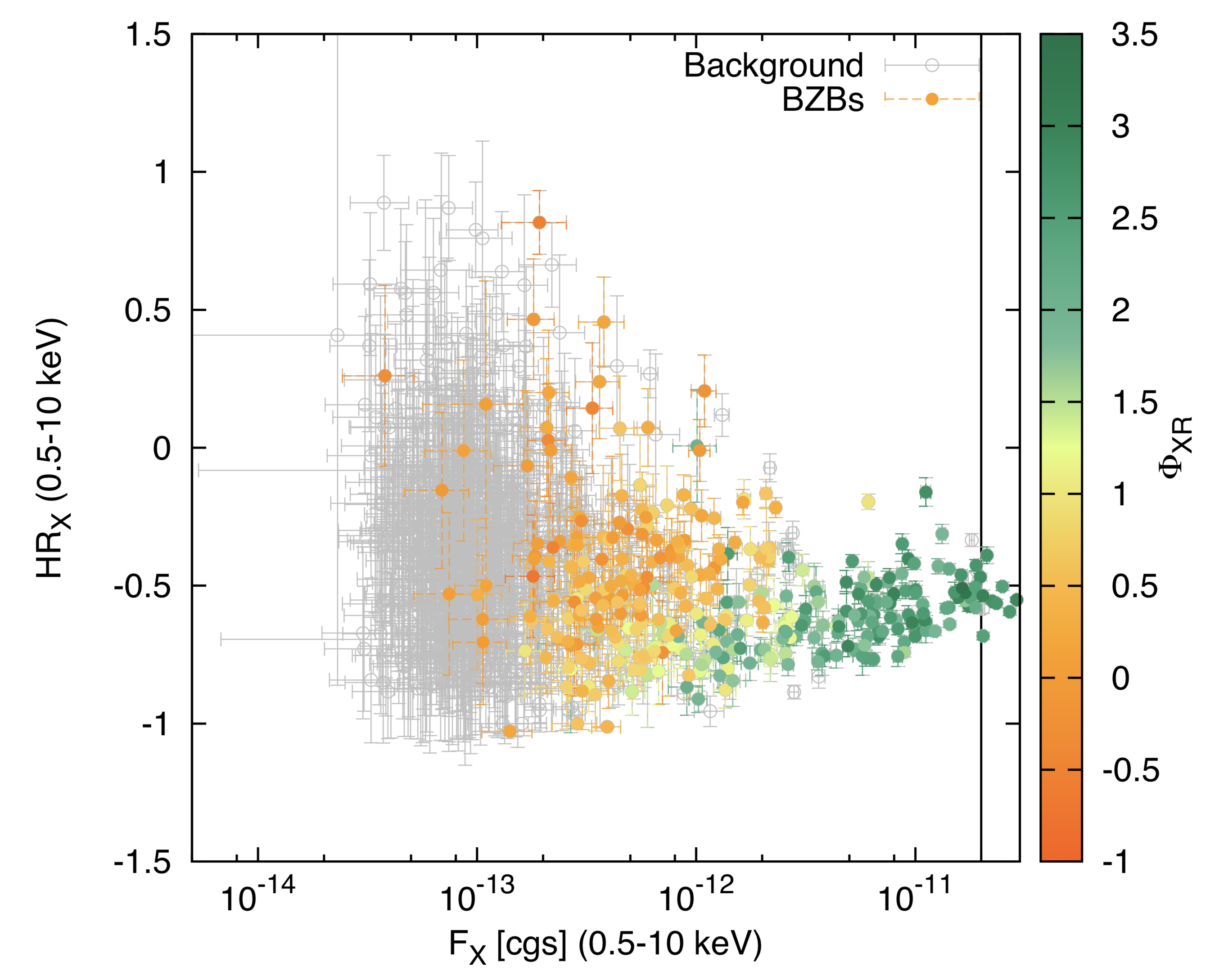}
\caption{$\rm{F}_{\rm{X}}$ in the 0.5-10 keV band vs. $\rm{HR}_{\rm{X}}$ for all BZBs in the selected sample. Background or foreground X-ray sources are marked in gray, and BZB classified as HBL and as LBL are marked in green and orange, respectively. The solid black line separates BZBs with pile-up, which show $\rm{F}_{\rm{X}}>2.0\times10^{-11}\,\rm{erg}\,\rm{cm}^{-2}\,\rm{s}^{-1}$.}
\end{figure}

\begin{figure}[ht!]
\centering 
\includegraphics[height=7.2cm,width=8.4cm,angle=0]{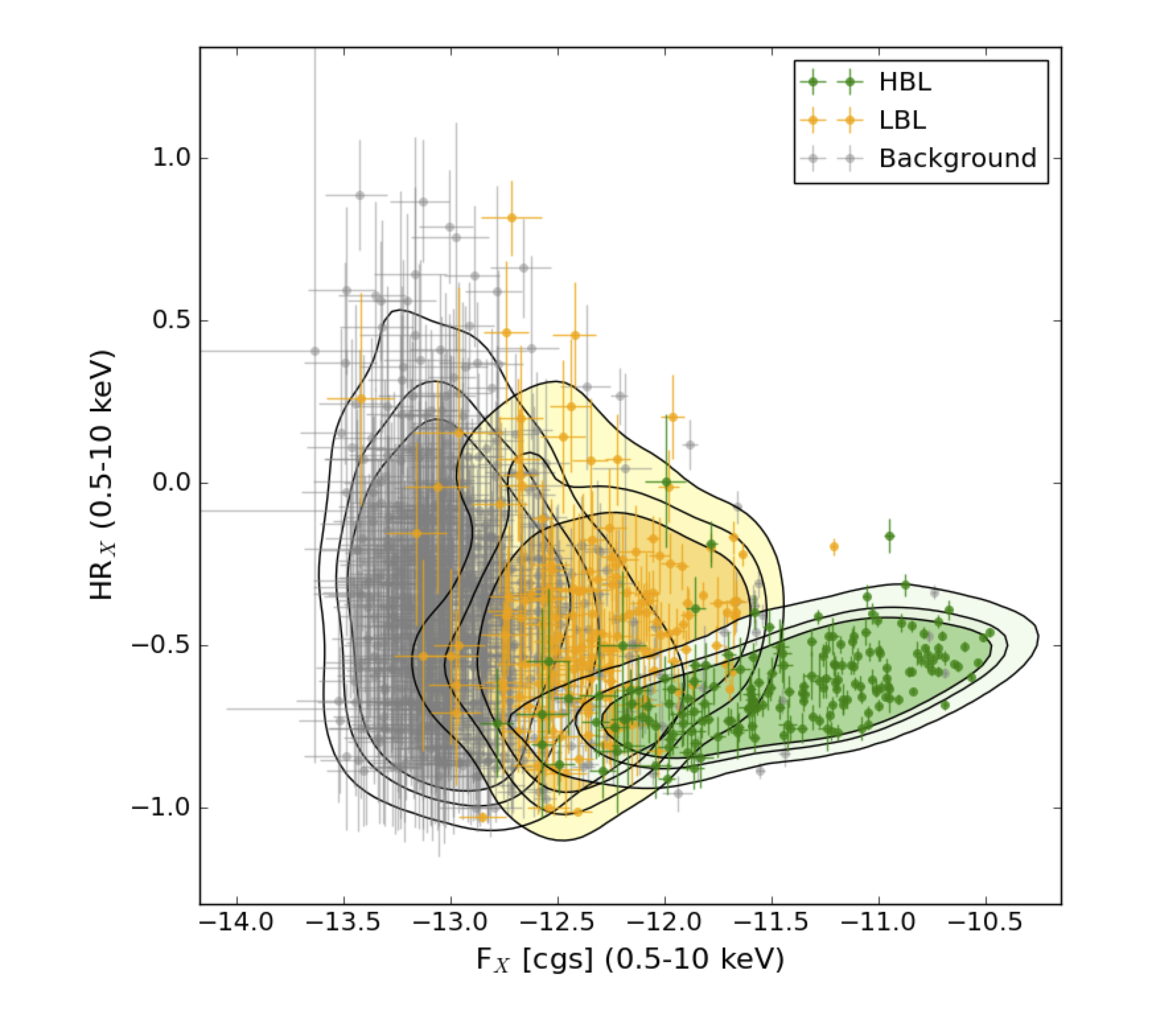}
\caption{$\rm{F}_{\rm{X}}$ in the 0.5-10 keV band vs. $\rm{HR}_{\rm{X}}$ for all BZBs in the selected sample, same as in Figure 5. We show the 70\%, 80\%, and 90\% isodensity contours as obtained from a KDE analysis for background or foreground sources in gray, HBLs in green, and LBLs in orange.}
\end{figure}

\begin{figure}[ht!]
\centering \includegraphics[height=7.6cm,width=8.4cm,angle=0]{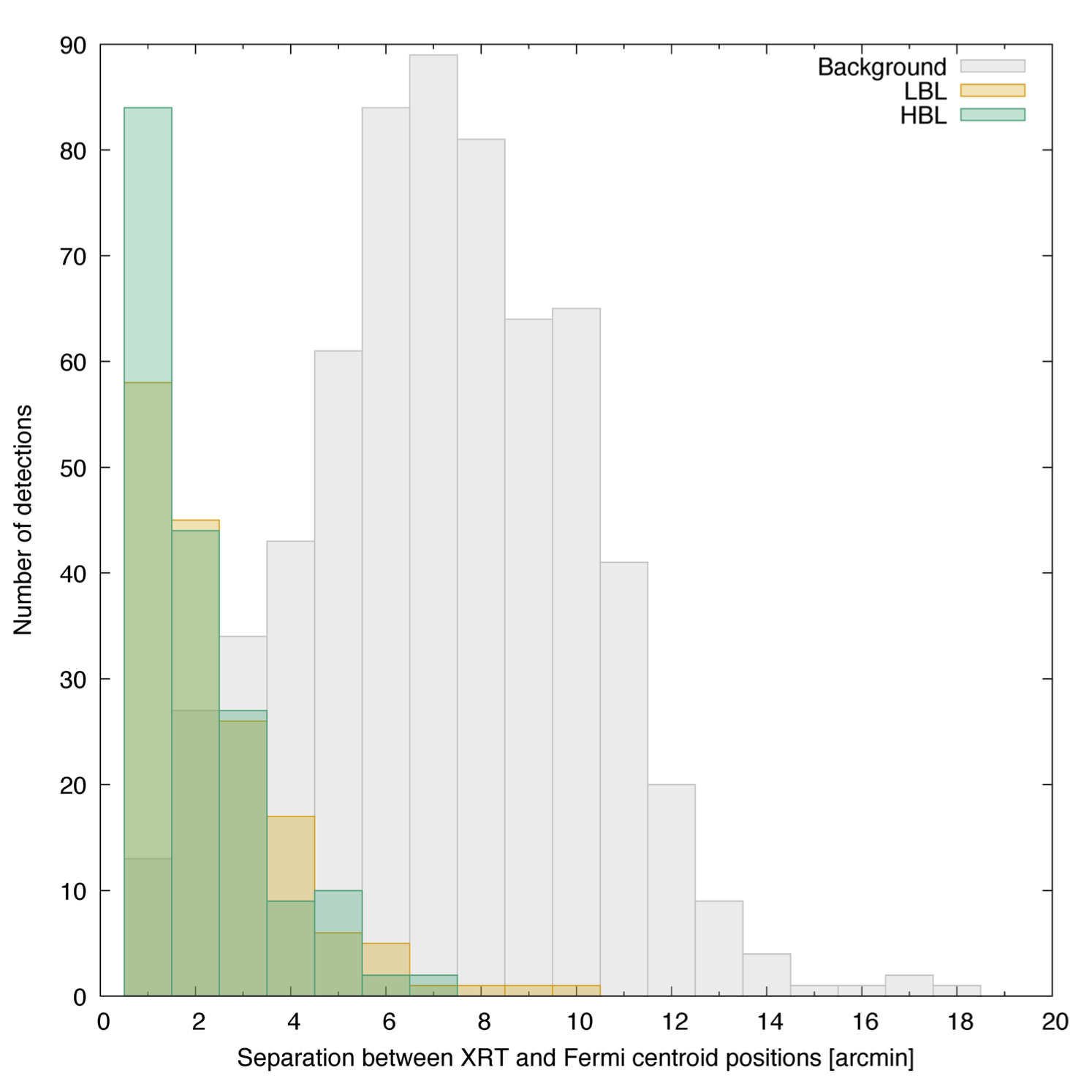}
\caption{Angular separation between the X-ray and $\gamma$-ray position for all BZBs, in arcminutes. As in Fig. 5, BZBs classified as HBLs and LBLs are shown in green and orange, respectively, while background or foreground X-ray sources are shown in gray.}
\end{figure}

In Figure 8 we report $\rm{F}_{\rm{X}}$ versus the WISE mid-infrared magnitude at 12 $\mu$m (in the upper panel), and versus the $\gamma$-ray energy flux in the 100 MeV - 100 GeV band, taken from the 3FGL catalog (in the upper panel). The 12 $\mu$m magnitude is the least affected by Galactic extinction but is still more sensitive than the W4 magnitude at 22 $\mu$m \citep{DAbrusco12}. No clear trend is visible in either panel of Figure 8. This is expected because while $\gamma$-rays are dominated by the IC SED component, X-rays are due to synchrotron emission in HBLs but might be a combination of synchrotron and IC components in LBLs \citep{Bondi01,Massaro08b}. No clear trend is visible between $\rm{F}_{\rm{X}}$ and the $\gamma$-ray energy flux. However, we highlight that because 96\% $\gamma$-ray BZBs have a counterpart in the X-rays, this provides a direct link between the BZB emission in these two energy ranges. We recall that HBLs are on average an order of magnitude brighter in X-rays than LBLs. The average $\rm{F}_{\rm{X}}$ value is $(4.5\pm3.3)\times10^{-12}\,\rm{erg}\,\rm{cm}^{-2}\,\rm{s}^{-1}$ for HBLs, and $(4.5\pm1.9)\times10^{-13}\,\rm{erg}\,\rm{cm}^{-2}\,\rm{s}^{-1}$ for LBLs. On the other hand, LBLs are on average 0.6 magnitudes brighter in mid-infrared than HBLs. 

\begin{figure}[ht!]
\centering \includegraphics[height=8.2cm,width=7.8cm,angle=0]{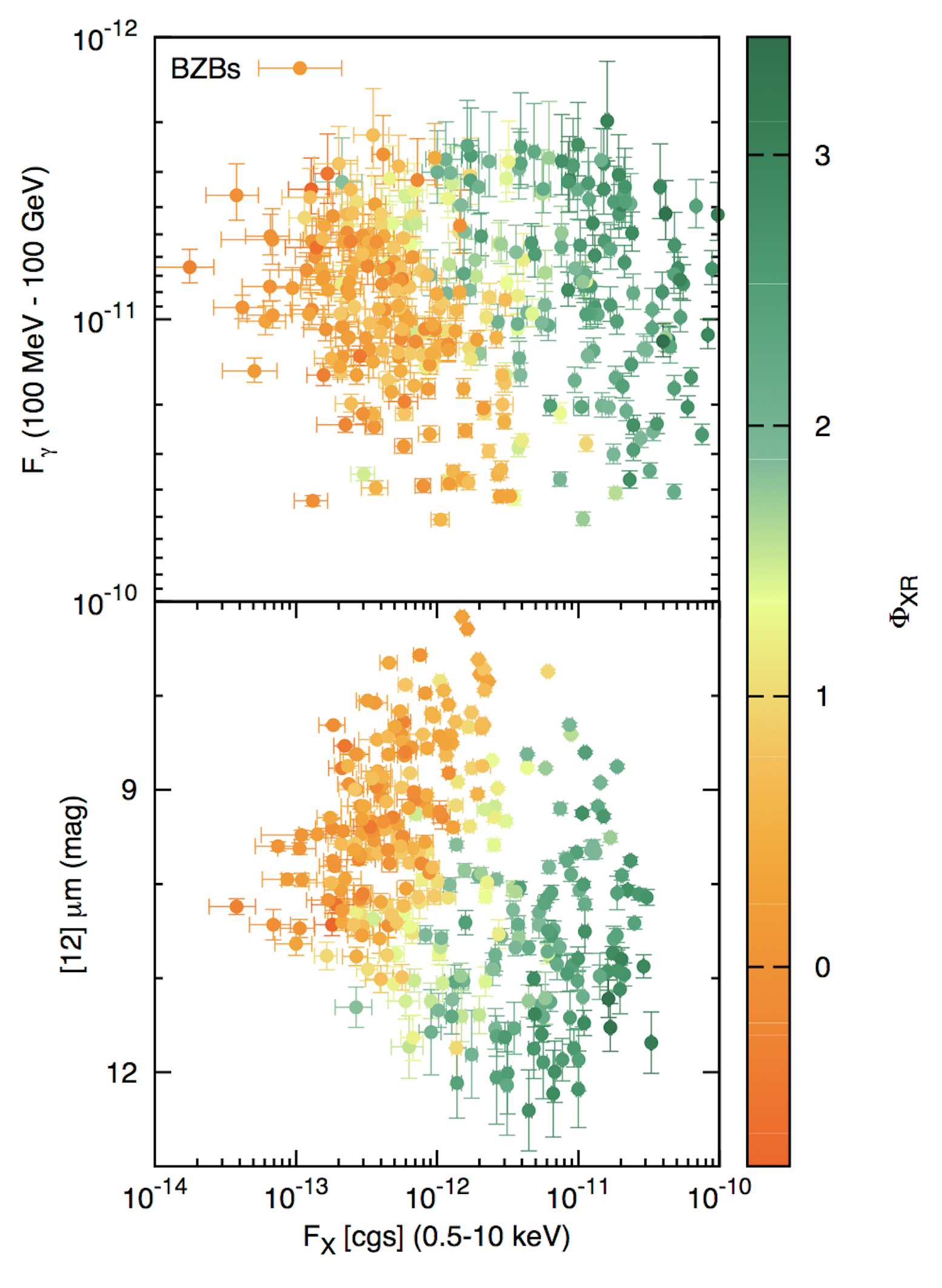}
\caption{Lower panel: $\rm{F}_{\rm{X}}$ in the 0.5-10 keV band vs. the WISE magnitude at 12 $\mu$m for all selected BZBs. Upper panel: $\rm{F}_{\rm{X}}$ in the 0.5-10 keV band vs. the $\rm{F}_{\gamma}$ in the 100 MeV - 100 GeV band, collected from the 3FGL. BL Lac objects classified as HBLs and LBLs are shown in green and orange, respectively.}
\end{figure}

\begin{figure}[ht!]
\centering \includegraphics[height=10.2cm,width=7.8cm,angle=0]{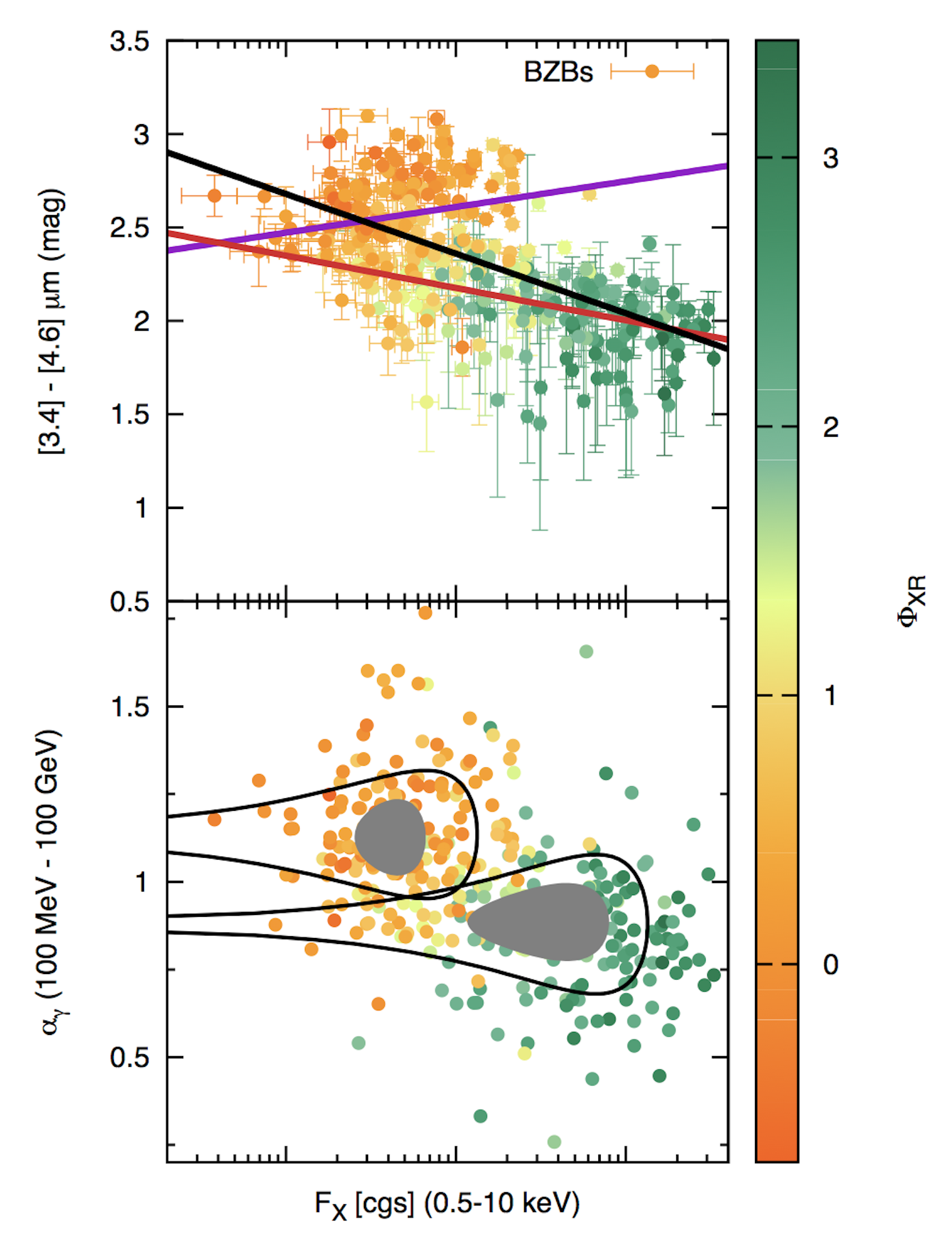}
\caption{Upper panel: $\rm{F}_{\rm{X}}$ in the 0.5-10 keV band vs. the [4.6]-[12] $\mu$m mid-infrared color. Selected BZBs are classified as HBLs and LBLs and are marked in green and orange, respectively. Dashed lines indicate regression lines for LBLs (purple), HBLs (red), and the whole sample (black). Lower panel: $\rm{F}_{\rm{X}}$ in the 0.5-10 keV band vs. the $\alpha_{\gamma}$ as reported in the 3FGL. Solid black lines mark the area within one standard deviation centered on the mean, and solid gray shows the area computed with the median absolute deviation centered on the median value.}
\end{figure}

The distinction between HBLs and LBLs becomes clearer when we compare their X-ray, $\gamma$-ray, and mid-infrared spectral shapes. In Figure 9 we show $\rm{F}_{\rm{X}}$ versus the mid-infrared color computed with WISE magnitudes at 4.6$\mu$m and 12$\mu$m (upper panel) and versus the $\gamma$-ray spectral index reported in the 3FGL (lower panel). HBLs are brighter in X-rays and harder in $\gamma$-rays than LBLs \citep{Ghisellini10,Ghisellini17}. In the lower panel of Figure 9 we also plot the average and median values of these parameters for both samples, together with their standard deviation and median absolute deviation, to highlight the clear distinction between the two subclasses. Although we were unable to establish a clear trend between fluxes or spectral shapes of BZBs in the X-ray and $\gamma$-ray band, we proved that there is a link between the emission in these two energy ranges given by the Fermi BZB high detection rate in the Swift/XRT observations analyzed here. 

This is different from what was discovered between radio, mid-infrared, and $\gamma$-ray emission for this class of AGNs, but it is crucial to support ongoing and future X-ray campaigns searching for blazars within sample of UGSs as well as for spectroscopic follow-up observations of X-ray sources that lie within the positional uncertainty region of unassociated Fermi sources.

The previous distinction improves significantly when the mid-infrared color is used, mainly because it marks the difference in the synchrotron component between HBLs and LBLs better. We found a correlation between $\rm{F}_{\rm{X}}$ and the [4.6]-[12]$\mu$m WISE color index, with a slope of $\sim -0.32$ and a p-chance of $\rm{p}<1\times10^{-7}$ for the whole sample of BZBs (plotted as the black line in Figure 9). This is due to the shifting of the whole synchrotron component toward higher energies because X-rays and infrared trace two different sides of the synchrotron peak. The trend is clearer for HBLs than for LBLs possibly because the IC component in the SEDs of the latter subclass might contribute. The p-chance of both correlations is $\rm{p}\sim3.7\times10^{-7}$ for HBLs (plotted in red) and $\rm{p}\sim0.019$ for LBLs (plotted in purple), indicating that the second is not statistically significant. In almost all previous WISE and X-ray combined investigations \citep[see, e.g.,][]{Maselli13}, the lack of uniform X-ray datasets did not allow a comparison of the behavior of HBLs and LBLs, as done here. 

In Figure 10 we show $\rm{HR}_{\rm{X}}$ versus the mid-infrared color and versus the $\gamma$-ray spectral index. We note that HBLs tend to cover a small range of X-ray hardness ratios; they appear to have softer X-ray spectra than LBLs. The average value for the $\rm{HR}_{\rm{X}}$ is $-0.63\pm0.09$ for HBLs and $-0.46\pm0.16$ for LBLs. The opposite occurs for the $\gamma$-ray spectral index and mid-infrared colors, where HBLs show a wider range of values than LBLs.

\begin{figure}[ht!]
\centering \includegraphics[height=10.2cm,width=7.8cm,angle=0]{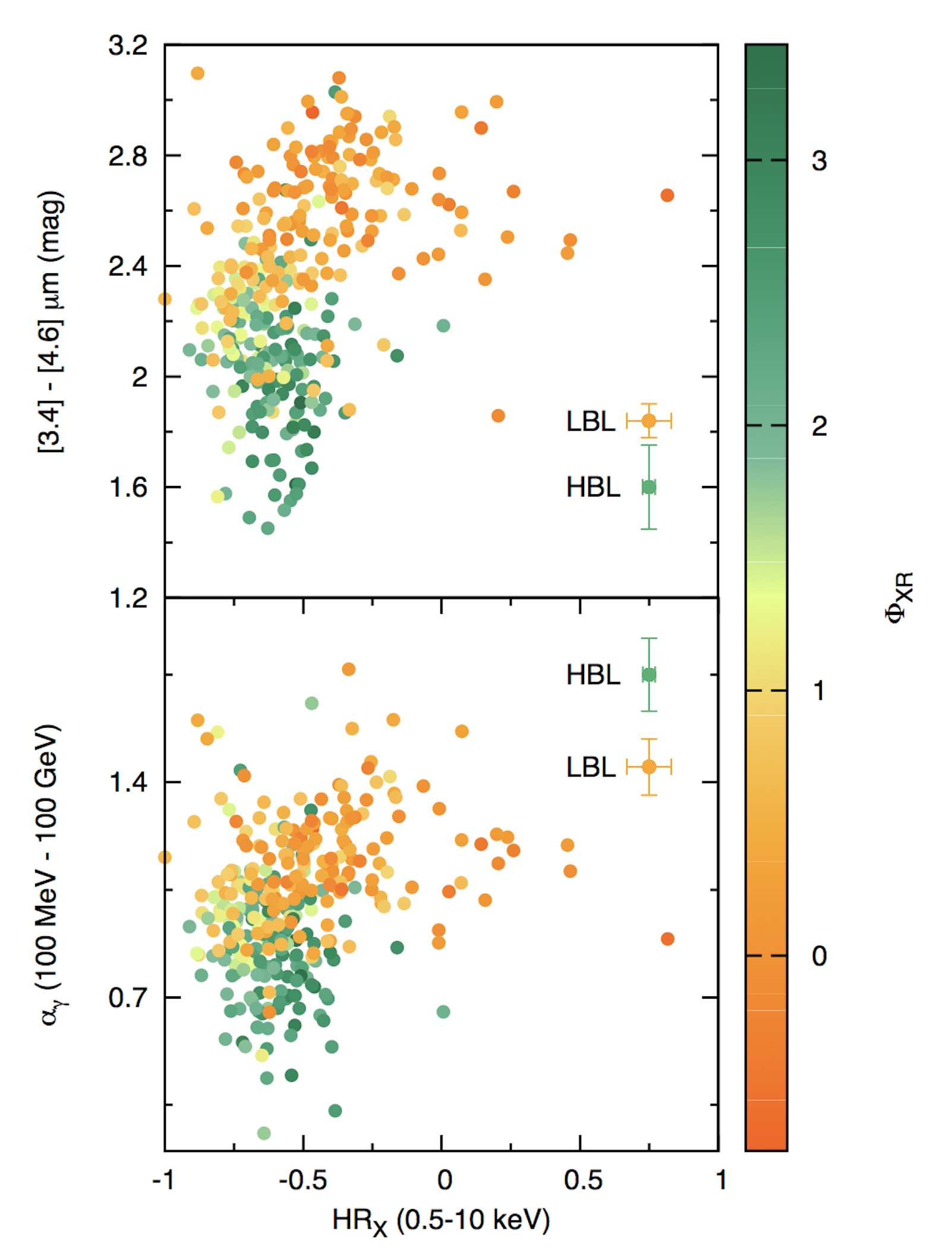}
\caption{Comparison of the spectral shape for all the BZBs in our sample. LBLs are shown in orange and HBLs in green. Upper panel: $\rm{HR}_{\rm{X}}$ in the 0.5-10 keV band vs. the [4.6]-[12] $\mu$m mid-infrared color. Lower panel: $\rm{HR}_{\rm{X}}$ in the 0.5-10 keV band vs. $\alpha_{\gamma}$. In both panels the average uncertainties on the parameters are shown as crosses in the top right corner.}
\end{figure}

Finally, for 77 BZBs a redshift estimate was available in Roma-BZCat, allowing us to compute their X-ray luminosities ($\rm{L}_{\rm{X}}$), as shown in Figure 11. HBLs tend to be more luminous than LBLs in the X-ray band. This might be due to the redshift difference, which on average is 0.283 for HBLs and 0.443 for LBLs. However, we also note that their corresponding $\gamma$-ray luminosities ($\rm{L}_{\gamma}$) are indeed quite similar for the same sample.

\begin{figure}[ht!]
\centering \includegraphics[scale=0.45]{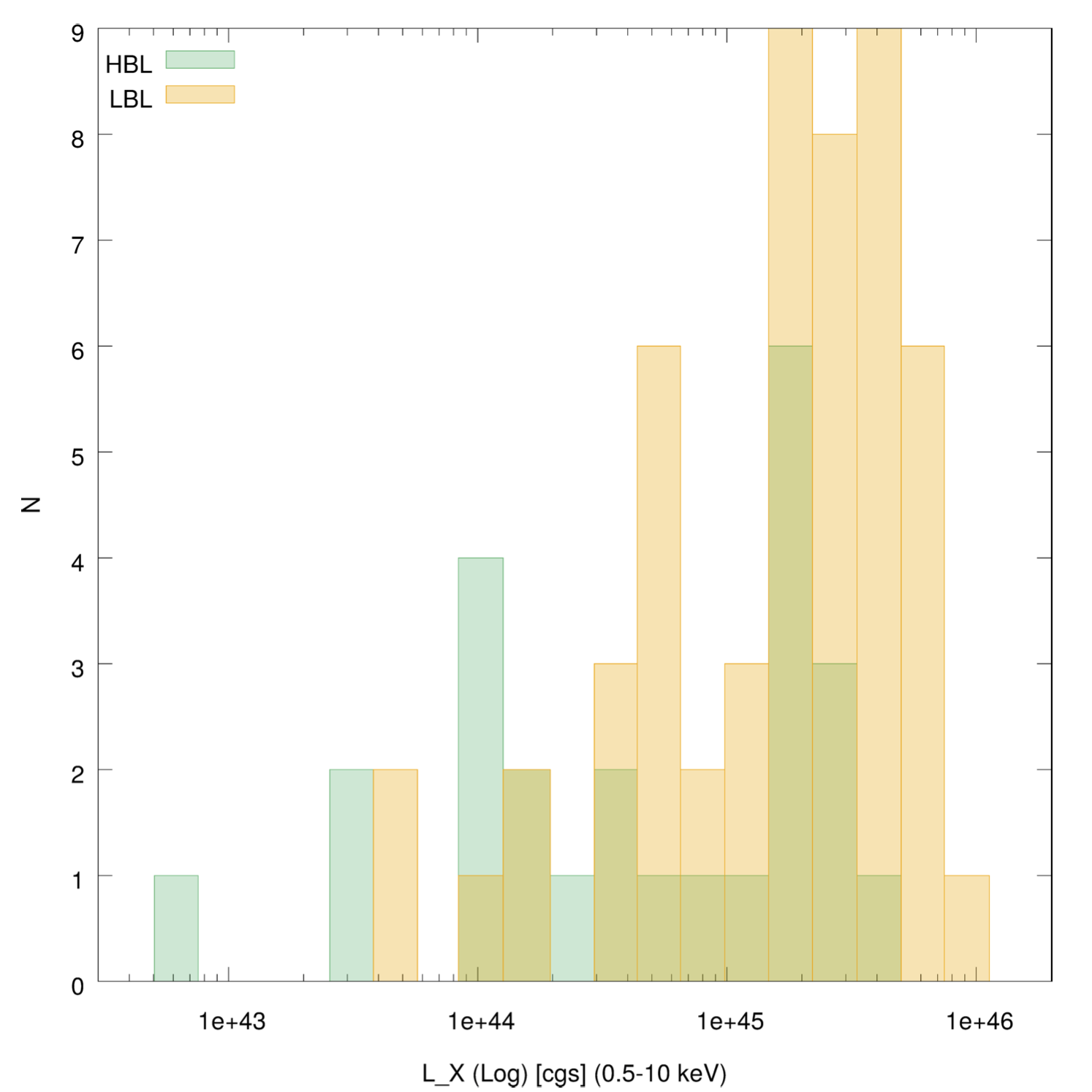}
\caption{Distribution of $\rm{L}_{\rm{X}}$ in the 0.5-10 keV band for the subsample of 77 BZBs with a well-determined redshift estimate reported in the Roma-BZCat \citep{Massaro15}. As in previous 
figures, LBLs are in orange and HBLs in green.}
\end{figure}

\section{Theoretical interpretation}

The theoretical connection between synchrotron and IC processes in the SSC scenario has previosly been extensively investigated \citep[see, e.g.,][and references therein]{Dermer95,Bloom96,Mastichiadis97,Dermer02,Massaro06,Tramacere11}. However, possible observable connections between X-ray and $\gamma$-ray emissions in BZBs are not yet fully exploited, certainly not on a statistical sample as we analyzed here. 

As previously stated, while we carried out this investigation, we found that HBLs tend to be brighter in the X-rays than LBLs and cover a wider range of fluxes. On the other hand, LBLs cover a wide range of $\rm{HR}_{\rm{X}}$, being generally harder than HBLs in the X-rays. These results suggest that on average HBLs do not vary their X-ray spectral shape significanlty, but only their total integrated power, and the opposite holds for LBLs. 

According to the SSC scenario, we could interpret this as follows. For LBLs a decrease in synchrotron emission in the X-rays could be balanced by an increase in IC component. This implies a certain balance in the total X-ray flux, but a noticeable change in the spectral shape (i.e., X-ray hardness ratio $\rm{HR}_{\rm{X}}$). On the other hand, the fact that for HBLs, $\rm{HR}_{\rm{X}}$ is restricted to a narrow range of values implies that the position of the synchrotron peak remains constant on average, and that we only observe the high-energy tail of their synchrotron component in X-rays.

We expect that the HBL behavior in X-rays is consistent with the theoretical scenario proposed in \citet{Massaro11a} on the basis of the acceleration mechanism proposed by \citet{Cavaliere80}. Thus, assuming that the beaming factor of HBL jets does not vary significantly, as the total number of emitting particles, and following \citet{Paggi09a}, we state that the frequency of the synchrotron peak scales as $\nu_{\rm{s}}\propto\gamma_{3\rm{p}}^2\rm{B}$, where $\rm{B}$ is the average magnetic field in the emission region and $\gamma_{3\rm{p}}$ is the peak of the $\gamma\,n(\gamma)$, with $n(\gamma)$ the particle energy distribution \citep[see][for details]{Tramacere07,Tramacere11}. 
For HBLs, $\nu_{\rm{s}}$ remains constant on average, implying that $\gamma_{3\rm{p}}^2\rm{B}=\rm{const}$. Assuming that the IC emission occurs in the Thomson regime, we expect that the SED peak frequency of the high-energy component scales as $\nu_{\rm{ic}} \propto \gamma_{3\rm{p}}\,\nu_{\rm{s}}$, thus as $\gamma_{3\rm{p}}^2$ under the circumstances previously described. On the other hand, the peak height of the IC component being in general proportional to $\gamma_{3\rm{p}}^4\,\rm{B}^2$ does not show significant changes if particle energy distribution and magnetic field are the main driver of spectral variations \citep{Paggi11}.

According to \citet{Blandford77} and \citet{Cavaliere02}, the BZB luminosity should be limited by the Blandford-Znajek (BZ) limit \citep{Ghosh97,Tchekhovskoy09}, defined as $\rm{L}_{\rm{BZ}}\leq 8\times10^{45}\times\left(\frac{\rm{M_{\rm{BH}}}}{10^{9}\,\rm{M_{\rm{Sun}}}}\right)\,\rm{erg}\,\rm{s}^{-1}$, where $\rm{M}_{\rm{BH}}$ is the central black hole mass, and $\rm{M}_{\rm{Sun}}$ the mass of the Sun. This condition applies only if the output can be ascribed solely to the black hole (i.e., in ``dry'' sources), which is a useful benchmark to analyze whether accretion is negligible in BZBs \citep{Paggi09b}.

There is evidence, however, that BZBs might be emitting more than what can be described within this benchmark alone \citep[see, e.g.,][]{Tavecchio16}. In presence of significant current accretion, we expect the jet launched from the central black hole to be more powerful  \citep{Ghisellini14}, and thus yielding higher observed luminosities that ultimately exceed the BZ limit, which instead applies to sources that are only powered by rotation.

For our sample we found 27 black hole mass estimates, measured with different methods as reported by \citet{Woo05,Plotkin11,LeonTavares11,Sbarrato12,Shaw12,Xiong14} and \citet{Ghisellini15}. We used them to compute the ratio of the sum of $\rm{L}_{\rm{X}}+\rm{L}_{\gamma}$ to the BZ luminosity limit. $\rm{L}_{\rm{X}}$ was computed from our data in the 0.5-10 keV band, and $\rm{L}_{\gamma}$ was obtained in the 100 MeV to 100 GeV band as listed in 3FGL. We adopted $\rm{L}_{\rm{X}}+\rm{L}_{\gamma}$ as an estimate of the bolometric luminosity because at least one of them is close to the SED peak. Thus, we expect the ratio to be below 1 if the BZ limit is applicable. In Figure 12 we plot this ratio versus the position of the rest-frame synchrotron peak taken from the 3FGL catalog.

\begin{figure}[ht!]
\centering \includegraphics[scale=0.45]{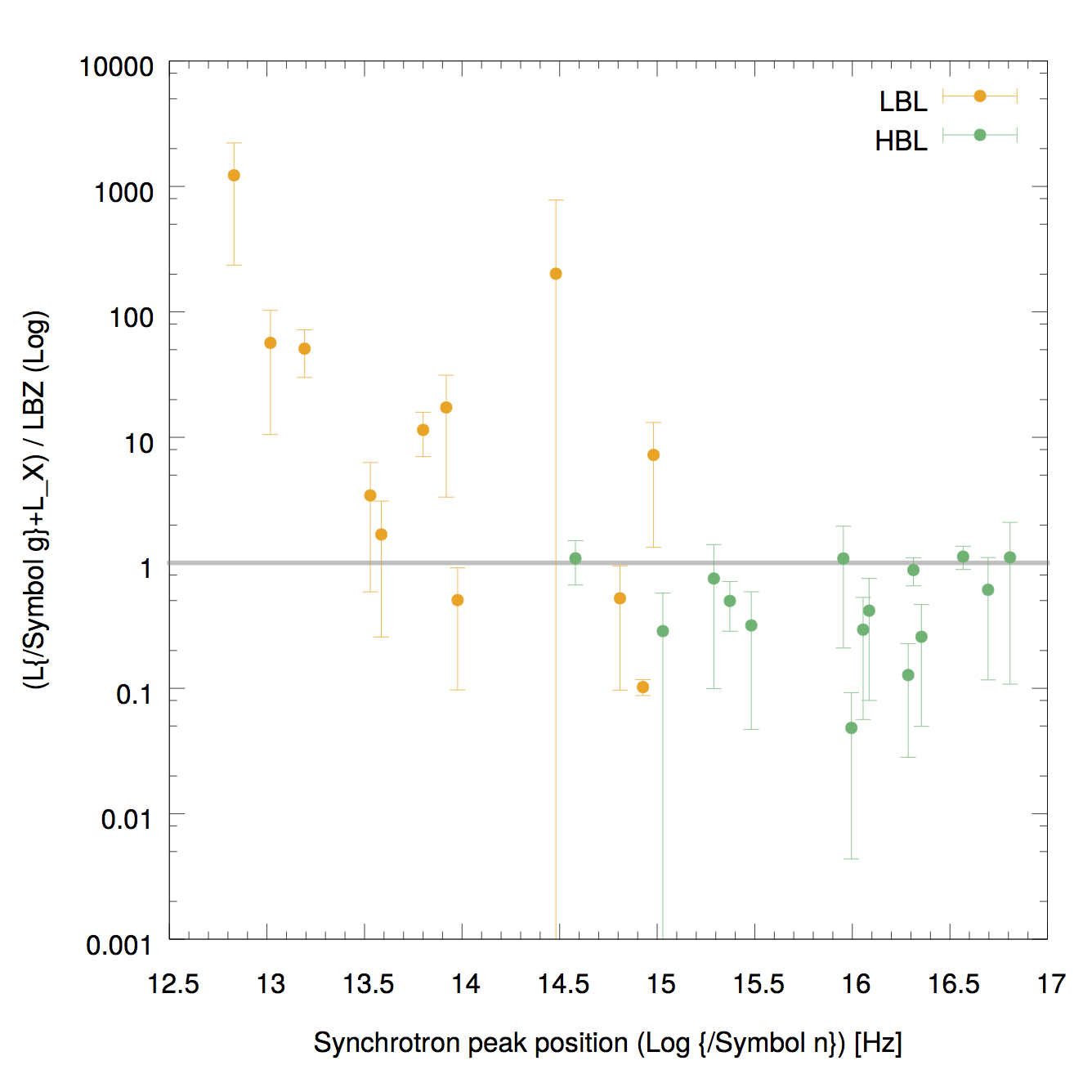}
\caption{Synchrotron peak frequency vs. the ratio between the estimate of the bolometric luminosity and the BZ luminosity, computed as reported in Section 5. As in previous figures, LBLs are in orange and HBLs in green. The gray dashed line indicates a luminosity ratio of 1, below which “dry” BZBs should lie.}
\end{figure}

It is clear that all HBLs are still below the BZ limit, which is expected because they are the less bolometrically luminous type of blazar \citep{Sambruna96}. In the LBL case, however, 8 out of 12 are above the BZ limit. Although the subsample is small, this is an indication that the BZ benchmark does not apply cleanly to the LBL scenario, as it does for HBLs.

An explanation for this discrepancy between the BZ benchmark and the observed LBL luminosities might be that these eight sources are probably not 'dry' sources as HBLs. Thus their broadband emission is contaminated by IC scattering of seed photons arising from surrounding gas and/or the accretion disk \citep{Abdo15,Arsioli18}. This is also consistent with the fact that LBLs are more similar to BZQs, and in these cases, the accretion component is important. In particular, for all these cases, $\rm{L}_{\gamma}$ is more than one order of magnitude higher than the X-ray luminosity, which is a strong indication that emission in these sources is strongly affected by an external Compton process \citep{Arsioli18}.

\section{Summary and conclusions}
\label{sec:conclusions}
We characterized the X-ray properties of \,\emph{Fermi} BZBs, searching for a possible connection between X-rays and  $\gamma$-raysthat could motivate follow-up X-ray observations to discover new blazar-like counterparts of UGSs \citep{Stroh13,Paggi13,Massaro15c}. To achieve our goal, we built a sample of 351 BZBs listed by \,\emph{Fermi} that were observed by the \emph{Swift}/XRT telescope in photon-counting mode, with exposure times between 1 and 20 ks, and collected up to December 2018.

\begin{enumerate}
    \item Of the 351 \,\emph{Fermi} BZBs that were observed by XRT for more than 1ks, 96\% (337)\ were  detected with an S/N greater than 3 (mean S/N ratio of 25). We obtained accurate positions, counts, count rates, fluxes, and images for all of them in the soft X-ray band between 0.5 and 10 keV.
    \item The BZBs are on average brighter in X-rays than other X-ray emitting sources that lie outside of the galactic plane ($|b|>10^{\circ}$). When they are classified as HBLs and LBLs, HBLs are generally brighter in X-ray flux and cover a wider range of flux values than LBLs.
    \item On the other hand, the spectral shape of LBLs in X-rays is harder and less uniform than that of HBLs. This might be due to a more variable spectral shape in the X-ray band: for HBLs, the X-ray emission is dominated by the synchrotron process, while for LBLs the IC component is non-negligible.
    \item We proved the direct link between the X-ray and $\gamma$-ray emission in BZBs, as supported by the high detection rate (96\%) of their X-ray counterparts. This rate would remain unchanged even with shorter exposure times (5 ks).
    \item The $\rm{F}_{\rm{X}}$ and mid-infrared color [4.6]-[12]$\mu$m ($p<1\times10^{-7}$) are correlated. X-ray brighter BZBs are bluer in the mid-infrared.
    \item For a subsample of 77 sources we obtained redshift values that allowed us to compute distances and luminosities. We note that HBLs are bolometrically less luminous than LBLs.
    \item For 27 sources out of these 77 with a determined redshift, we also searched for values of the central black hole mass in the literature. All HBLs are below their BZ limit. However, some LBLs exceed this limit. We speculate that this indicates that LBLs may be accreting more gas than HBLs, meaning that the 'dry' hypothesis would be valid only in the latter, not in the former.
\end{enumerate}

\section{Acknowledgements}
E. J. Marchesini would like to thank Roc\'io I. P\'aez and M. Victoria Reynaldi for useful discussions on this work. This work is supported by the ``Departments of Excellence 2018 - 2022’’ Grant awarded by the Italian Ministry of Education, University and Research (MIUR) (L. 232/2016). This research has made use of resources provided by the Compagnia di San Paolo for the grant awarded on the BLENV project (S1618\_L1\_MASF\_01) and by the Ministry of Education, Universities and Research for the grant MASF\_FFABR\_17\_01. F.M. acknowledges financial contribution from the agreement ASI-INAF n.2017-14-H.0 A.P. acknowledges financial support from the Consorzio Interuniversitario per la fisica Spaziale (CIFS) under the agreement related to the grant MASF\_CONTR\_FIN\_18\_02. This research has made use of data obtained from the high- energy Astrophysics Science Archive Research Center (HEASARC) provided by NASA’s Goddard Space Flight Center. This work is part of a project that has received funding from the European Union's Horizon 2020 Research and Innovation Programme under the Marie S\l{}odowska-Curie Grant Agreement No. 664931. Part of this work is based on the NVSS (NRAO VLA Sky Survey). The National Radio Astronomy Observatory is operated by Associated Universities, Inc., under contract with the National Science Foundation. The Molonglo Observatory site manager, Duncan Campbell-Wilson, and the staff, Jeff Webb, Michael White, and John Barry, are responsible for the smooth operation of the Molonglo Observatory Synthesis Telescope (MOST) and the day-to-day observing program of SUMSS. SUMSS is dedicated to Michael Large, whose expertise and vision made the project possible. The MOST is operated by the School of Physics with the support of the Australian Research Council and the Science Foundation for Physics within the University of Sydney. This publication makes use of data products from the Wide-field Infrared Survey Explorer, which is a joint project of the University of California, Los Angeles, and the Jet Propulsion Laboratory/California Institute of Technology, funded by the National Aeronautics and Space Administration. TOPCAT\footnote{http://www.star.bris.ac.uk/m~bt/topcat/} \citep{Taylor05} and STILTS \citep{Taylor06} were used for the preparation and manipulation of the images and the tabular data.

\section{Appendix}

In this section we include complementary figures. In Figure 13 we show the X-ray full-band flux versus the magnitudes at 3.4 and 4.6 $\mu$m from the AllWISE catalog in the upper and lower panel, respectively. They do not differ from the results in the main text, as shown in Figure 10. In the same way, we show in Figure 14 the comparison between the X-ray full-band flux and the color index [3.4]-[4.6], which is comparable to Figure 11. Finally, in Figure 15 we show the X-ray hardness ratio versus the color index [3.4]-[4.6], which follows the same trend as discussed for Figure 10.

\begin{figure}
\centering \includegraphics[height=10.2cm,width=7.8cm,angle=0]{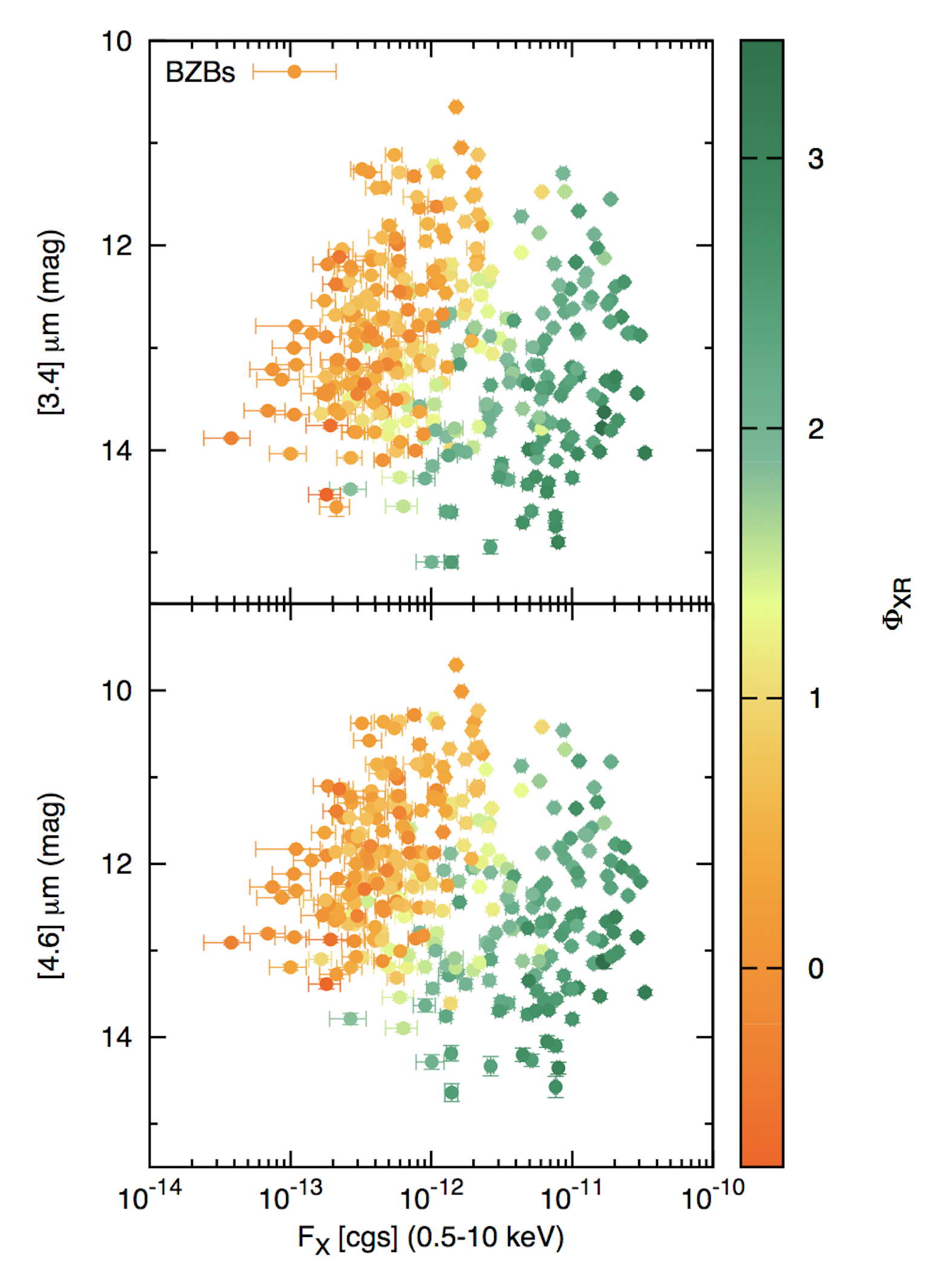}
\caption{$\rm{F}_{\rm{X}}$ in the 0.5-10 keV band vs. the WISE magnitude estimate at a nominal wavelength of 3.4 $\mu$m (upper panel) and vs. that evaluated at 4.6 $\mu$m (lower panel). BZBs classified as LBL are shown in orange and HBLs 
are marked in green.}
\end{figure}

\begin{figure}
\centering \includegraphics[height=6.2cm,width=8.2cm,angle=0]{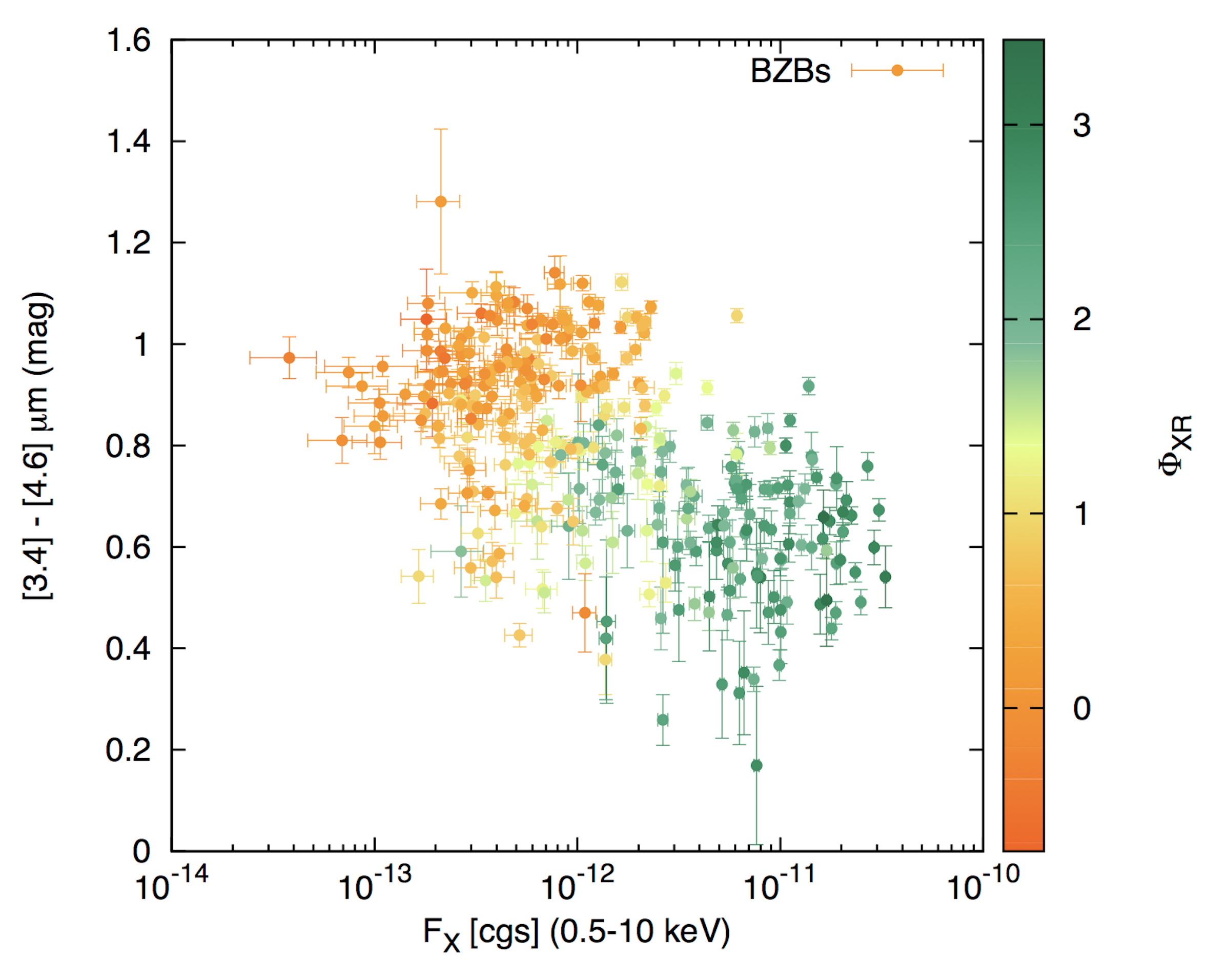}
\caption{$\rm{F}_{\rm{X}}$ in the 0.5-10 keV band vs. the [3.4]-[4.6] $\mu$m mid-infrared color.}
\end{figure}

\begin{figure}
\centering \includegraphics[height=6.2cm,width=8.2cm,angle=0]{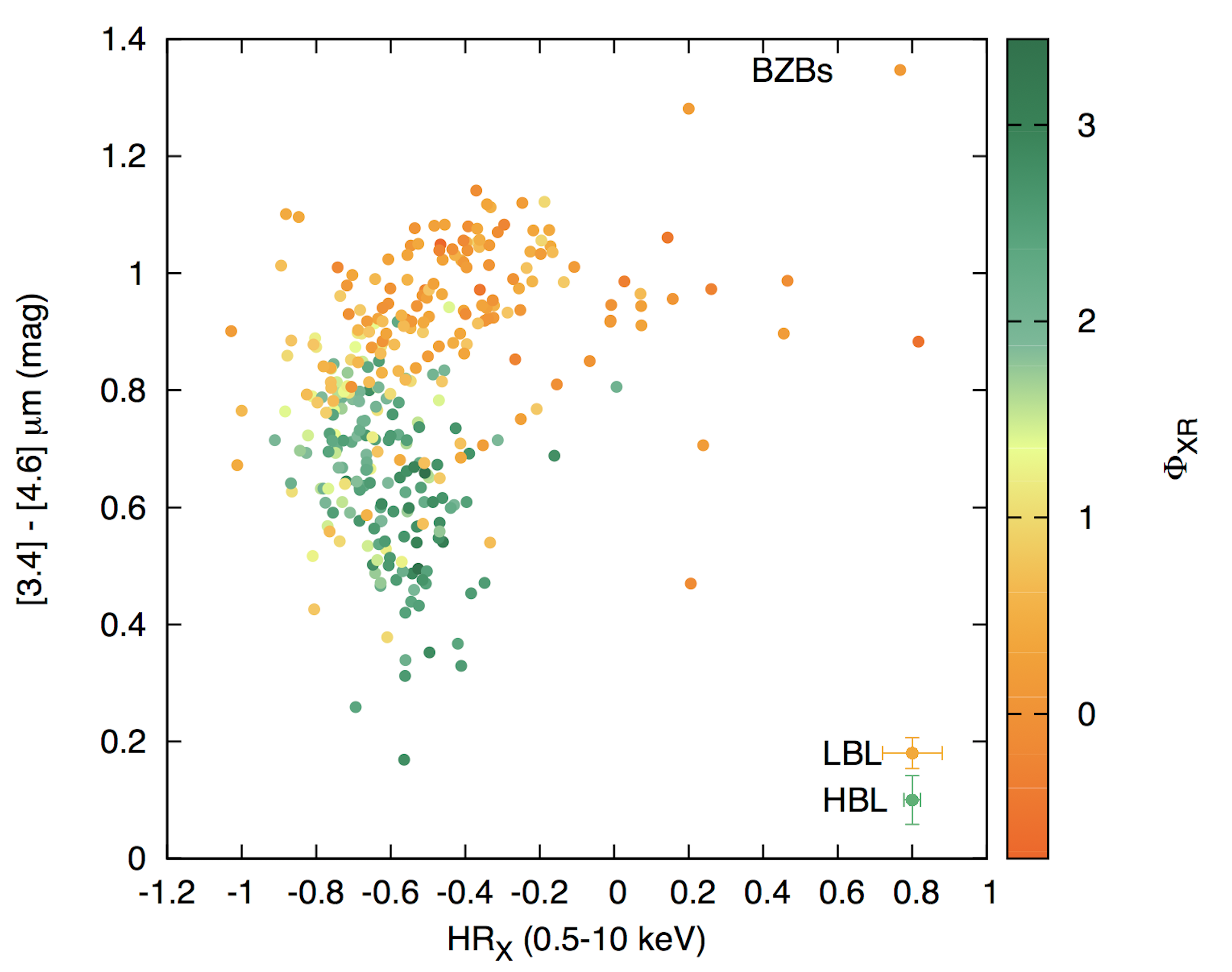}
\caption{$\rm{HR}_{\rm{X}}$ in the 0.5-10 keV band vs. the [3.4]-[4.6] $\mu$m. Average uncertainties on these two parameters are shown as crosses in the bottom right corner.}
\end{figure}

\newpage
\bibliography{Biblio} 

\begin{thebibliography}{145}
\expandafter\ifx\csname natexlab\endcsname\relax\def\natexlab#1{#1}\fi

\bibitem[{{Abdo} {et~al.}(2015){Abdo}, {Ackermann}, {Ajello}, {Allafort},
  {Amin}, {Baldini}, {Barbiellini}, {Bastieri}, {Bechtol}, \&
  {Bellazzini}}]{Abdo15}
{Abdo}, A.~A., {Ackermann}, M., {Ajello}, M., {et~al.} 2015, \apj, 799, 143

\bibitem[{{Acero} {et~al.}(2015){Acero}, {Ackermann}, {Ajello}, {Albert},
  {Atwood}, {Axelsson}, {Baldini}, {Ballet}, {Barbiellini}, {Bastieri},
  {Belfiore}, {Bellazzini}, {Bissaldi}, {Blandford}, {Bloom}, {Bogart},
  {Bonino}, {Bottacini}, {Bregeon}, {Britto}, {Bruel}, {Buehler}, {Burnett},
  {Buson}, {Caliandro}, {Cameron}, {Caputo}, {Caragiulo}, {Caraveo},
  {Casandjian}, {Cavazzuti}, {Charles}, {Chaves}, {Chekhtman}, {Cheung},
  {Chiang}, {Chiaro}, {Ciprini}, {Claus}, {Cohen-Tanugi}, {Cominsky}, {Conrad},
  {Cutini}, {D'Ammando}, {de Angelis}, {DeKlotz}, {de Palma}, {Desiante},
  {Digel}, {Di Venere}, {Drell}, {Dubois}, {Dumora}, {Favuzzi}, {Fegan},
  {Ferrara}, {Finke}, {Franckowiak}, {Fukazawa}, {Funk}, {Fusco}, {Gargano},
  {Gasparrini}, {Giebels}, {Giglietto}, {Giommi}, {Giordano}, {Giroletti},
  {Glanzman}, {Godfrey}, {Grenier}, {Grondin}, {Grove}, {Guillemot}, {Guiriec},
  {Hadasch}, {Harding}, {Hays}, {Hewitt}, {Hill}, {Horan}, {Iafrate}, {Jogler},
  {J{\'o}hannesson}, {Johnson}, {Johnson}, {Johnson}, {Johnson}, {Kamae},
  {Kataoka}, {Katsuta}, {Kuss}, {La Mura}, {Landriu}, {Larsson}, {Latronico},
  {Lemoine-Goumard}, {Li}, {Li}, {Longo}, {Loparco}, {Lott}, {Lovellette},
  {Lubrano}, {Madejski}, {Massaro}, {Mayer}, {Mazziotta}, {McEnery},
  {Michelson}, {Mirabal}, {Mizuno}, {Moiseev}, {Mongelli}, {Monzani},
  {Morselli}, {Moskalenko}, {Murgia}, {Nuss}, {Ohno}, {Ohsugi}, {Omodei},
  {Orienti}, {Orlando}, {Ormes}, {Paneque}, {Panetta}, {Perkins},
  {Pesce-Rollins}, {Piron}, {Pivato}, {Porter}, {Racusin}, {Rando}, {Razzano},
  {Razzaque}, {Reimer}, {Reimer}, {Reposeur}, {Rochester}, {Romani},
  {Salvetti}, {S{\'a}nchez-Conde}, {Saz Parkinson}, {Schulz}, {Siskind},
  {Smith}, {Spada}, {Spandre}, {Spinelli}, {Stephens}, {Strong}, {Suson},
  {Takahashi}, {Takahashi}, {Tanaka}, {Thayer}, {Thayer}, {Thompson},
  {Tibaldo}, {Tibolla}, {Torres}, {Torresi}, {Tosti}, {Troja}, {Van Klaveren},
  {Vianello}, {Winer}, {Wood}, {Wood}, {Zimmer}, \& {Fermi-LAT
  Collaboration}}]{Acero15}
{Acero}, F., {Ackermann}, M., {Ajello}, M., {et~al.} 2015, \apjs, 218, 23

\bibitem[{{Ackermann} {et~al.}(2011){Ackermann}, {Ajello}, {Allafort},
  {Angelakis}, {Axelsson}, {Baldini}, {Ballet}, {Barbiellini}, {Bastieri},
  {Bellazzini}, {Berenji}, {Blandford}, {Bloom}, {Bonamente}, {Borgland},
  {Bouvier}, {Bregeon}, {Brez}, {Brigida}, {Bruel}, {Buehler}, {Buson},
  {Caliandro}, {Cameron}, {Cannon}, {Caraveo}, {Casandjian}, {Cavazzuti},
  {Cecchi}, {Charles}, {Chekhtman}, {Cheung}, {Ciprini}, {Claus},
  {Cohen-Tanugi}, {Cutini}, {de Palma}, {Dermer}, {Silva}, {Drell}, {Dubois},
  {Dumora}, {Escande}, {Favuzzi}, {Fegan}, {Focke}, {Fortin}, {Frailis},
  {Fuhrmann}, {Fukazawa}, {Fusco}, {Gargano}, {Gasparrini}, {Gehrels},
  {Giglietto}, {Giommi}, {Giordano}, {Giroletti}, {Glanzman}, {Godfrey},
  {Grandi}, {Grenier}, {Guiriec}, {Hadasch}, {Hayashida}, {Hays}, {Healey},
  {J{\'o}hannesson}, {Johnson}, {Kamae}, {Katagiri}, {Kataoka},
  {Kn{\"o}dlseder}, {Kuss}, {Lande}, {Lee}, {Longo}, {Loparco}, {Lott},
  {Lovellette}, {Lubrano}, {Makeev}, {Max-Moerbeck}, {Mazziotta}, {McEnery},
  {Mehault}, {Michelson}, {Mizuno}, {Monte}, {Monzani}, {Morselli},
  {Moskalenko}, {Murgia}, {Naumann-Godo}, {Nishino}, {Nolan}, {Norris}, {Nuss},
  {Ohsugi}, {Okumura}, {Omodei}, {Orlando}, {Ormes}, {Ozaki}, {Paneque},
  {Pavlidou}, {Pelassa}, {Pepe}, {Pesce-Rollins}, {Pierbattista}, {Piron},
  {Porter}, {Rain{\`o}}, {Razzano}, {Readhead}, {Reimer}, {Reimer}, {Richards},
  {Romani}, {Sadrozinski}, {Scargle}, {Sgr{\`o}}, {Siskind}, {Smith},
  {Spandre}, {Spinelli}, {Strickman}, {Suson}, {Takahashi}, {Tanaka}, {Taylor},
  {Thayer}, {Thayer}, {Thompson}, {Torres}, {Tosti}, {Tramacere}, {Troja},
  {Vandenbroucke}, {Vianello}, {Vitale}, {Waite}, {Wang}, {Winer}, {Wood},
  {Yang}, \& {Ziegler}}]{Ackermann11}
{Ackermann}, M., {Ajello}, M., {Allafort}, A., {et~al.} 2011, \apj, 741, 30

\bibitem[{{Ackermann} {et~al.}(2015){Ackermann}, {Ajello}, {Atwood}, {Baldini},
  {Ballet}, {Barbiellini}, {Bastieri}, {Becerra Gonzalez}, {Bellazzini},
  {Bissaldi}, {Blandford}, {Bloom}, {Bonino}, {Bottacini}, {Brandt}, {Bregeon},
  {Britto}, {Bruel}, {Buehler}, {Buson}, {Caliandro}, {Cameron}, {Caragiulo},
  {Caraveo}, {Carpenter}, {Casandjian}, {Cavazzuti}, {Cecchi}, {Charles},
  {Chekhtman}, {Cheung}, {Chiang}, {Chiaro}, {Ciprini}, {Claus},
  {Cohen-Tanugi}, {Cominsky}, {Conrad}, {Cutini}, {D'Abrusco}, {D'Ammando}, {de
  Angelis}, {Desiante}, {Digel}, {Di Venere}, {Drell}, {Favuzzi}, {Fegan},
  {Ferrara}, {Finke}, {Focke}, {Franckowiak}, {Fuhrmann}, {Fukazawa},
  {Furniss}, {Fusco}, {Gargano}, {Gasparrini}, {Giglietto}, {Giommi},
  {Giordano}, {Giroletti}, {Glanzman}, {Godfrey}, {Grenier}, {Grove},
  {Guiriec}, {Hewitt}, {Hill}, {Horan}, {Itoh}, {J{\'o}hannesson}, {Johnson},
  {Johnson}, {Kataoka}, {Kawano}, {Krauss}, {Kuss}, {La Mura}, {Larsson},
  {Latronico}, {Leto}, {Li}, {Li}, {Longo}, {Loparco}, {Lott}, {Lovellette},
  {Lubrano}, {Madejski}, {Mayer}, {Mazziotta}, {McEnery}, {Michelson},
  {Mizuno}, {Moiseev}, {Monzani}, {Morselli}, {Moskalenko}, {Murgia}, {Nuss},
  {Ohno}, {Ohsugi}, {Ojha}, {Omodei}, {Orienti}, {Orlando}, {Paggi}, {Paneque},
  {Perkins}, {Pesce-Rollins}, {Piron}, {Pivato}, {Porter}, {Rain{\`o}},
  {Rando}, {Razzano}, {Razzaque}, {Reimer}, {Reimer}, {Romani}, {Salvetti},
  {Schaal}, {Schinzel}, {Schulz}, {Sgr{\`o}}, {Siskind}, {Sokolovsky}, {Spada},
  {Spandre}, {Spinelli}, {Stawarz}, {Suson}, {Takahashi}, {Takahashi},
  {Tanaka}, {Thayer}, {Thayer}, {Tibaldo}, {Torres}, {Torresi}, {Tosti},
  {Troja}, {Uchiyama}, {Vianello}, {Winer}, {Wood}, \& {Zimmer}}]{Ackermann15}
{Ackermann}, M., {Ajello}, M., {Atwood}, W.~B., {et~al.} 2015, \apj, 810, 14

\bibitem[{{Aharonian} {et~al.}(2005){Aharonian}, {Akhperjanian}, {Aye},
  {Bazer-Bachi}, {Beilicke}, {Benbow}, {Berge}, {Berghaus}, {Bernl{\"o}hr},
  {Boisson}, {Bolz}, {Braun}, {Breitling}, {Brown}, {Bussons Gordo},
  {Chadwick}, {Chounet}, {Cornils}, {Costamante}, {Degrange},
  {Djannati-Ata{\"i}}, {O'C.~Drury}, {Dubus}, {Emmanoulopoulos}, {Espigat},
  {Feinstein}, {Fleury}, {Fontaine}, {Fuchs}, {Funk}, {Gallant}, {Giebels},
  {Gillessen}, {Glicenstein}, {Goret}, {Hadjichristidis}, {Hauser},
  {Heinzelmann}, {Henri}, {Hermann}, {Hinton}, {Hofmann}, {Holleran}, {Horns},
  {de Jager}, {Kh{\'e}lifi}, {Komin}, {Konopelko}, {Latham}, {Le Gallou},
  {Lemi{\`e}re}, {Lemoine-Goumard}, {Leroy}, {Lohse}, {Martineau-Huynh},
  {Marcowith}, {Masterson}, {McComb}, {de Naurois}, {Nolan}, {Noutsos},
  {Orford}, {Osborne}, {Ouchrif}, {Panter}, {Pelletier}, {Pita},
  {P{\"u}hlhofer}, {Punch}, {Raubenheimer}, {Raue}, {Raux}, {Rayner},
  {Redondo}, {Reimer}, {Reimer}, {Ripken}, {Rob}, {Rolland}, {Rowell},
  {Sahakian}, {Saug{\'e}}, {Schlenker}, {Schlickeiser}, {Schuster}, {Schwanke},
  {Siewert}, {Sol}, {Steenkamp}, {Stegmann}, {Tavernet}, {Terrier},
  {Th{\'e}oret}, {Tluczykont}, {Vasileiadis}, {Venter}, {Vincent}, {V{\"o}lk},
  \& {Wagner}}]{Aharonian05}
{Aharonian}, F., {Akhperjanian}, A.~G., {Aye}, K.-M., {et~al.} 2005, \aap, 436,
  L17

\bibitem[{{Albert} {et~al.}(2007){Albert}, {Aliu}, {Anderhub}, {Antoranz},
  {Armada}, {Baixeras}, {Barrio}, {Bartko}, {Bastieri}, {Becker}, {Bednarek},
  {Berger}, {Bigongiari}, {Biland}, {Bock}, {Bordas}, {Bosch-Ramon}, {Bretz},
  {Britvitch}, {Camara}, {Carmona}, {Chilingarian}, {Coarasa}, {Commichau},
  {Contreras}, {Cortina}, {Costado}, {Curtef}, {Danielyan}, {Dazzi}, {De
  Angelis}, {Delgado}, {de los Reyes}, {De Lotto}, {Domingo-Santamar{\'{\i}}a},
  {Dorner}, {Doro}, {Errando}, {Fagiolini}, {Ferenc}, {Fern{\'a}ndez}, {Firpo},
  {Flix}, {Fonseca}, {Font}, {Fuchs}, {Galante}, {Garc{\'{\i}}a-L{\'o}pez},
  {Garczarczyk}, {Gaug}, {Giller}, {Goebel}, {Hakobyan}, {Hayashida},
  {Hengstebeck}, {Herrero}, {H{\"o}hne}, {Hose}, {Hrupec}, {Hsu}, {Jacon},
  {Jogler}, {Kosyra}, {Kranich}, {Kritzer}, {Laille}, {Lindfors}, {Lombardi},
  {Longo}, {L{\'o}pez}, {L{\'o}pez}, {Lorenz}, {Majumdar}, {Maneva},
  {Mannheim}, {Mansutti}, {Mariotti}, {Mart{\'{\i}}nez}, {Mazin}, {Merck},
  {Meucci}, {Meyer}, {Miranda}, {Mirzoyan}, {Mizobuchi}, {Moralejo}, {Nieto},
  {Nilsson}, {Ninkovic}, {O{\~n}a-Wilhelmi}, {Otte}, {Oya}, {Paneque},
  {Panniello}, {Paoletti}, {Paredes}, {Pasanen}, {Pascoli}, {Pauss}, {Pegna},
  {Persic}, {Peruzzo}, {Piccioli}, {Prandini}, {Puchades}, {Raymers}, {Rhode},
  {Rib{\'o}}, {Rico}, {Rissi}, {Robert}, {R{\"u}gamer}, {Saggion}, {Saito},
  {S{\'a}nchez}, {Sartori}, {Scalzotto}, {Scapin}, {Schmitt}, {Schweizer},
  {Shayduk}, {Shinozaki}, {Shore}, {Sidro}, {Sillanp{\"a}{\"a}}, {Sobczynska},
  {Stamerra}, {Stark}, {Takalo}, {Tavecchio}, {Temnikov}, {Tescaro}, {Teshima},
  {Torres}, {Turini}, {Vankov}, {Vitale}, {Wagner}, {Wibig}, {Wittek},
  {Zandanel}, {Zanin}, \& {Zapatero}}]{Albert07}
{Albert}, J., {Aliu}, E., {Anderhub}, H., {et~al.} 2007, \apj, 669, 862

\bibitem[{{{\'A}lvarez Crespo} {et~al.}(2016{\natexlab{a}}){{\'A}lvarez
  Crespo}, {Masetti}, {Ricci}, {Landoni}, {Pati{\~n}o-{\'A}lvarez}, {Massaro},
  {D'Abrusco}, {Paggi}, {Chavushyan}, {Jim{\'e}nez-Bail{\'o}n}, {Torrealba},
  {Latronico}, {La Franca}, {Smith}, \& {Tosti}}]{Opt5}
{{\'A}lvarez Crespo}, N., {Masetti}, N., {Ricci}, F., {et~al.}
  2016{\natexlab{a}}, \aj, 151, 32

\bibitem[{{{\'A}lvarez Crespo} {et~al.}(2016{\natexlab{b}}){{\'A}lvarez
  Crespo}, {Massaro}, {D'Abrusco}, {Land oni}, {Masetti}, {Chavushyan},
  {Jim{\'e}nez-Bail{\'o}n}, {La Franca}, {Milisavljevic}, \&
  {Paggi}}]{AlvarezCrespo16}
{{\'A}lvarez Crespo}, N., {Massaro}, F., {D'Abrusco}, R., {et~al.}
  2016{\natexlab{b}}, \apss, 361, 316

\bibitem[{{{\'A}lvarez Crespo} {et~al.}(2016{\natexlab{c}}){{\'A}lvarez
  Crespo}, {Massaro}, {Milisavljevic}, {Landoni}, {Chavushyan},
  {Pati{\~n}o-{\'A}lvarez}, {Masetti}, {Jim{\'e}nez-Bail{\'o}n}, {Strader},
  {Chomiuk}, {Katagiri}, {Kagaya}, {Cheung}, {Paggi}, {D'Abrusco}, {Ricci}, {La
  Franca}, {Smith}, \& {Tosti}}]{Opt6}
{{\'A}lvarez Crespo}, N., {Massaro}, F., {Milisavljevic}, D., {et~al.}
  2016{\natexlab{c}}, \aj, 151, 95

\bibitem[{{Arsioli} \& {Chang}(2018)}]{Arsioli18}
{Arsioli}, B. \& {Chang}, Y.~L. 2018, \aap, 616, A63

\bibitem[{{Banerjee} {et~al.}(2019){Banerjee}, {Joshi}, {Majumdar},
  {Williamson}, {Jorstad}, \& {Marscher}}]{Banerjee19}
{Banerjee}, B., {Joshi}, M., {Majumdar}, P., {et~al.} 2019, \mnras, 487, 845

\bibitem[{{Blandford} \& {Rees}(1978)}]{BlandfordRees78}
{Blandford}, R.~D. \& {Rees}, M.~J. 1978, in BL Lac Objects, ed. A.~M. {Wolfe},
  328--341

\bibitem[{{Blandford} \& {Znajek}(1977)}]{Blandford77}
{Blandford}, R.~D. \& {Znajek}, R.~L. 1977, \mnras, 179, 433

\bibitem[{{Bloom}(2008)}]{Bloom08}
{Bloom}, S.~D. 2008, \aj, 136, 1533

\bibitem[{{Bloom} \& {Marscher}(1996)}]{Bloom96}
{Bloom}, S.~D. \& {Marscher}, A.~P. 1996, \apj, 461, 657

\bibitem[{{Bondi} {et~al.}(2001){Bondi}, {March{\~a}}, {Dallacasa}, \&
  {Stanghellini}}]{Bondi01}
{Bondi}, M., {March{\~a}}, M.~J.~M., {Dallacasa}, D., \& {Stanghellini}, C.
  2001, \mnras, 325, 1109

\bibitem[{{B{\"o}ttcher} {et~al.}(2007){B{\"o}ttcher}, {Basu}, {Joshi},
  {Villata}, {Arai}, {Aryan}, {Asfandiyarov}, {Bach}, {Bachev}, {Berduygin},
  {Blaek}, {Buemi}, {Castro-Tirado}, {De Ugarte Postigo}, {Frasca}, {Fuhrmann},
  {Hagen-Thorn}, {Henson}, {Hovatta}, {Hudec}, {Ibrahimov}, {Ishii},
  {Ivanidze}, {Jel{\'{\i}}nek}, {Kamada}, {Kapanadze}, {Katsuura}, {Kotaka},
  {Kovalev}, {Kovalev}, {Kub{\'a}nek}, {Kurosaki}, {Kurtanidze},
  {L{\"a}hteenm{\"a}ki}, {Lanteri}, {Larionov}, {Larionova}, {Lee}, {Leto},
  {Lindfors}, {Marilli}, {Marshall}, {Miller}, {Mingaliev}, {Mirabal},
  {Mizoguchi}, {Nakamura}, {Nieppola}, {Nikolashvili}, {Nilsson}, {Nishiyama},
  {Ohlert}, {Osterman}, {Pak}, {Pasanen}, {Peters}, {Pursimo}, {Raiteri},
  {Robertson}, {Robertson}, {Ryle}, {Sadakane}, {Sadun}, {Sigua}, {Sohn},
  {Strigachev}, {Sumitomo}, {Takalo}, {Tamesue}, {Tanaka}, {Thorstensen},
  {Tosti}, {Trigilio}, {Umana}, {Vennes}, {Vitek}, {Volvach}, {Webb},
  {Yamanaka}, \& {Yim}}]{Bottcher07}
{B{\"o}ttcher}, M., {Basu}, S., {Joshi}, M., {et~al.} 2007, \apj, 670, 968

\bibitem[{{Brinkmann} {et~al.}(2005){Brinkmann}, {Papadakis}, {Raeth},
  {Mimica}, \& {Haberl}}]{Brinkmann05}
{Brinkmann}, W., {Papadakis}, I.~E., {Raeth}, C., {Mimica}, P., \& {Haberl}, F.
  2005, \aap, 443, 397

\bibitem[{{Bruni} {et~al.}(2018){Bruni}, {Panessa}, {Ghisellini}, {Chavushyan},
  {Pe{\~n}a-Herazo}, {Hern{\'a}ndez-Garc{\'{\i}}a}, {Bazzano}, {Ubertini}, \&
  {Kraus}}]{Bruni18}
{Bruni}, G., {Panessa}, F., {Ghisellini}, G., {et~al.} 2018, \apjl, 854, L23

\bibitem[{{Capalbi} {et~al.}(2005){Capalbi}, {Perri}, {Saija}, \&
  {Tamburelli}}]{Capalbi05}
{Capalbi}, M., {Perri}, M., {Saija}, B., \& {Tamburelli}, F. 2005, ASI Science
  Data Center, 1

\bibitem[{{Carini} {et~al.}(1992){Carini}, {Miller}, {Noble}, \&
  {Goodrich}}]{Carini92}
{Carini}, M.~T., {Miller}, H.~R., {Noble}, J.~C., \& {Goodrich}, B.~D. 1992,
  \aj, 104, 15

\bibitem[{{Cavaliere} \& {D'Elia}(2002)}]{Cavaliere02}
{Cavaliere}, A. \& {D'Elia}, V. 2002, \apj, 571, 226

\bibitem[{{Cavaliere} \& {Morrison}(1980)}]{Cavaliere80}
{Cavaliere}, A. \& {Morrison}, P. 1980, \apj, 238, L63

\bibitem[{{Ciaramella} {et~al.}(2004){Ciaramella}, {Bongardo}, {Aller},
  {Aller}, {De Zotti}, {L{\"a}hteenmaki}, {Longo}, {Milano}, {Tagliaferri},
  {Ter{\"a}sranta}, {Tornikoski}, \& {Urpo}}]{Ciaramella04}
{Ciaramella}, A., {Bongardo}, C., {Aller}, H.~D., {et~al.} 2004, \aap, 419, 485

\bibitem[{{Condon} {et~al.}(1998){Condon}, {Cotton}, {Greisen}, {Yin},
  {Perley}, {Taylor}, \& {Broderick}}]{Condon98}
{Condon}, J.~J., {Cotton}, W.~D., {Greisen}, E.~W., {et~al.} 1998, \aj, 115,
  1693

\bibitem[{{Cutini} {et~al.}(2014){Cutini}, {Ciprini}, {Orienti}, {Tramacere},
  {D'Ammando}, {Verrecchia}, {Polenta}, {Carrasco}, {D'Elia}, {Giommi},
  {Gonz{\'a}lez-Nuevo}, {Grandi}, {Harrison}, {Hays}, {Larsson},
  {L{\"a}hteenm{\"a}ki}, {Le{\'o}n-Tavares}, {L{\'o}pez-Caniego}, {Natoli},
  {Ojha}, {Partridge}, {Porras}, {Reyes}, {Recillas}, \& {Torresi}}]{Cutini14}
{Cutini}, S., {Ciprini}, S., {Orienti}, M., {et~al.} 2014, \mnras, 445, 4316

\bibitem[{{D'Abrusco} {et~al.}(2009){D'Abrusco}, {Longo}, \&
  {Walton}}]{DAbrusco09}
{D'Abrusco}, R., {Longo}, G., \& {Walton}, N.~A. 2009, \mnras, 396, 223

\bibitem[{{D'Abrusco} {et~al.}(2012){D'Abrusco}, {Massaro}, {Ajello},
  {Grindlay}, {Smith}, \& {Tosti}}]{DAbrusco12}
{D'Abrusco}, R., {Massaro}, F., {Ajello}, M., {et~al.} 2012, \apj, 748, 68

\bibitem[{{D'Abrusco} {et~al.}(2013){D'Abrusco}, {Massaro}, {Paggi}, {Masetti},
  {Tosti}, {Giroletti}, \& {Smith}}]{DAbrusco13}
{D'Abrusco}, R., {Massaro}, F., {Paggi}, A., {et~al.} 2013, \apjs, 206, 12

\bibitem[{{D'Abrusco} {et~al.}(2014){D'Abrusco}, {Massaro}, {Paggi}, {Smith},
  {Masetti}, {Landoni}, \& {Tosti}}]{DAbrusco14}
{D'Abrusco}, R., {Massaro}, F., {Paggi}, A., {et~al.} 2014, \apjs, 215, 14

\bibitem[{{D'Elia} {et~al.}(2013){D'Elia}, {Perri}, {Puccetti}, {Capalbi},
  {Giommi}, {Burrows}, {Campana}, {Tagliaferri}, {Cusumano}, {Evans},
  {Gehrels}, {Kennea}, {Moretti}, {Nousek}, {Osborne}, {Romano}, \&
  {Stratta}}]{DElia13}
{D'Elia}, V., {Perri}, M., {Puccetti}, S., {et~al.} 2013, \aap, 551, A142

\bibitem[{{Dermer}(1995)}]{Dermer95}
{Dermer}, C.~D. 1995, \apj, 446, L63

\bibitem[{{Dermer} \& {Schlickeiser}(2002)}]{Dermer02}
{Dermer}, C.~D. \& {Schlickeiser}, R. 2002, \apj, 575, 667

\bibitem[{{D{\textquoteright}Abrusco}
  {et~al.}(2019){D{\textquoteright}Abrusco}, {{\'A}lvarez Crespo}, {Massaro},
  {Campana}, {Chavushyan}, {Landoni}, {La Franca}, {Masetti}, {Milisavljevic},
  \& {Paggi}}]{DAbrusco19}
{D{\textquoteright}Abrusco}, R., {{\'A}lvarez Crespo}, N., {Massaro}, F.,
  {et~al.} 2019, \apjs, 242, 4

\bibitem[{{Dunkley} {et~al.}(2009){Dunkley}, {Komatsu}, {Nolta}, {Spergel},
  {Larson}, {Hinshaw}, {Page}, {Bennett}, {Gold}, {Jarosik}, {Weiland},
  {Halpern}, {Hill}, {Kogut}, {Limon}, {Meyer}, {Tucker}, {Wollack}, \&
  {Wright}}]{Dunkley09}
{Dunkley}, J., {Komatsu}, E., {Nolta}, M.~R., {et~al.} 2009, \apjs, 180, 306

\bibitem[{{Evans} {et~al.}(2014){Evans}, {Osborne}, {Beardmore}, {Page},
  {Willingale}, {Mountford}, {Pagani}, {Burrows}, {Kennea}, {Perri},
  {Tagliaferri}, \& {Gehrels}}]{Evans14}
{Evans}, P.~A., {Osborne}, J.~P., {Beardmore}, A.~P., {et~al.} 2014, \apjs,
  210, 8

\bibitem[{{Falcone}(2013)}]{Falcone13}
{Falcone}, A. 2013, in AAS/High Energy Astrophysics Division, Vol.~13, AAS/High
  Energy Astrophysics Division \#13, 114.07

\bibitem[{{Falcone} {et~al.}(2014){Falcone}, {Stroh}, \& {Pryal}}]{Falcone14}
{Falcone}, A., {Stroh}, M., \& {Pryal}, M. 2014, in American Astronomical
  Society Meeting Abstracts, Vol. 223, American Astronomical Society Meeting
  Abstracts \#223, 301.05

\bibitem[{{Fichtel} {et~al.}(1993){Fichtel}, {Bertsch}, {Dingus}, {Hartman},
  {Hunter}, {Kanbach}, {Kniffen}, {Kwok}, {Lin}, \& {Mattox}}]{Fichtel93}
{Fichtel}, C.~E., {Bertsch}, D.~L., {Dingus}, B., {et~al.} 1993, Advances in
  Space Research, 13, 637

\bibitem[{{Finke} {et~al.}(2008){Finke}, {Dermer}, \& {B{\"o}ttcher}}]{Finke08}
{Finke}, J.~D., {Dermer}, C.~D., \& {B{\"o}ttcher}, M. 2008, \apj, 686, 181

\bibitem[{{Ghirlanda} {et~al.}(2010){Ghirlanda}, {Ghisellini}, {Tavecchio}, \&
  {Foschini}}]{Ghirlanda10}
{Ghirlanda}, G., {Ghisellini}, G., {Tavecchio}, F., \& {Foschini}, L. 2010,
  \mnras, 407, 791

\bibitem[{{Ghirlanda} {et~al.}(2011){Ghirlanda}, {Ghisellini}, {Tavecchio},
  {Foschini}, \& {Bonnoli}}]{Ghirlanda11}
{Ghirlanda}, G., {Ghisellini}, G., {Tavecchio}, F., {Foschini}, L., \&
  {Bonnoli}, G. 2011, \mnras, 413, 852

\bibitem[{{Ghisellini} {et~al.}(1985){Ghisellini}, {Maraschi}, \&
  {Treves}}]{Ghisellini85}
{Ghisellini}, G., {Maraschi}, L., \& {Treves}, A. 1985, \aap, 146, 204

\bibitem[{{Ghisellini} {et~al.}(2017){Ghisellini}, {Righi}, {Costamante}, \&
  {Tavecchio}}]{Ghisellini17}
{Ghisellini}, G., {Righi}, C., {Costamante}, L., \& {Tavecchio}, F. 2017,
  \mnras, 469, 255

\bibitem[{{Ghisellini} \& {Tavecchio}(2015)}]{Ghisellini15}
{Ghisellini}, G. \& {Tavecchio}, F. 2015, \mnras, 448, 1060

\bibitem[{{Ghisellini} {et~al.}(2010){Ghisellini}, {Tavecchio}, {Foschini},
  {Ghirland a}, {Maraschi}, \& {Celotti}}]{Ghisellini10}
{Ghisellini}, G., {Tavecchio}, F., {Foschini}, L., {et~al.} 2010, \mnras, 402,
  497

\bibitem[{{Ghisellini} {et~al.}(2014){Ghisellini}, {Tavecchio}, {Maraschi},
  {Celotti}, \& {Sbarrato}}]{Ghisellini14}
{Ghisellini}, G., {Tavecchio}, F., {Maraschi}, L., {Celotti}, A., \&
  {Sbarrato}, T. 2014, \nat, 515, 376

\bibitem[{{Ghosh} \& {Abramowicz}(1997)}]{Ghosh97}
{Ghosh}, P. \& {Abramowicz}, M.~A. 1997, \mnras, 292, 887

\bibitem[{{Giannios} {et~al.}(2009){Giannios}, {Uzdensky}, \&
  {Begelman}}]{Giannios09}
{Giannios}, D., {Uzdensky}, D.~A., \& {Begelman}, M.~C. 2009, \mnras, 395, L29

\bibitem[{{Giommi} {et~al.}(1990){Giommi}, {Barr}, {Garilli}, {Maccagni}, \&
  {Pollock}}]{Giommi90}
{Giommi}, P., {Barr}, P., {Garilli}, B., {Maccagni}, D., \& {Pollock}, A.~M.~T.
  1990, \apj, 356, 432

\bibitem[{{Giroletti} {et~al.}(2016){Giroletti}, {Massaro}, {D'Abrusco},
  {Lico}, {Burlon}, {Hurley-Walker}, {Johnston-Hollitt}, {Morgan}, {Pavlidou},
  {Bell}, {Bernardi}, {Bhat}, {Bowman}, {Briggs}, {Cappallo}, {Corey},
  {Deshpande}, {Ewall-Rice}, {Emrich}, {Gaensler}, {Goeke}, {Greenhill},
  {Hazelton}, {Hindson}, {Kaplan}, {Kasper}, {Kratzenberg}, {Feng}, {Jacobs},
  {Kudryavtseva}, {Lenc}, {Lonsdale}, {Lynch}, {McKinley}, {McWhirter},
  {Mitchell}, {Morales}, {Morgan}, {Oberoi}, {Offringa}, {Ord}, {Pindor},
  {Prabu}, {Procopio}, {Riding}, {Rogers}, {Roshi}, {Udaya Shankar}, {Srivani},
  {Subrahmanyan}, {Tingay}, {Waterson}, {Wayth}, {Webster}, {Whitney},
  {Williams}, \& {Williams}}]{Giroletti16}
{Giroletti}, M., {Massaro}, F., {D'Abrusco}, R., {et~al.} 2016, \aap, 588, A141

\bibitem[{{Hartman} {et~al.}(1999){Hartman}, {Bertsch}, {Bloom}, {Chen},
  {Deines-Jones}, {Esposito}, {Fichtel}, {Friedlander}, {Hunter}, \&
  {McDonald}}]{Hartman99}
{Hartman}, R.~C., {Bertsch}, D.~L., {Bloom}, S.~D., {et~al.} 1999, \apjs, 123,
  79

\bibitem[{{Hartman} {et~al.}(2001){Hartman}, {B{\"o}ttcher}, {Aldering},
  {Aller}, {Aller}, {Backman}, {Balonek}, {Bertsch}, {Bloom}, {Bock},
  {Boltwood}, {Carini}, {Collmar}, {De Francesco}, {Ferrara}, {Freudling},
  {Gear}, {Hall}, {Heidt}, {Hughes}, {Hunter}, {Jogee}, {Johnson}, {Kanbach},
  {Katajainen}, {Kidger}, {Kii}, {Koskimies}, {Kraus}, {Kubo}, {Kurtanidze},
  {Lanteri}, {Lawson}, {Lin}, {Lisenfeld}, {Madejski}, {Makino}, {Maraschi},
  {Marscher}, {McFarland}, {McHardy}, {Miller}, {Nikolashvili}, {Nilsson},
  {Noble}, {Nucciarelli}, {Ostorero}, {Pian}, {Pursimo}, {Raiteri}, {Reich},
  {Rekola}, {Richter}, {Robson}, {Sadun}, {Savolainen}, {Sillanp{\"a}{\"a}},
  {Smale}, {Sobrito}, {Sreekumar}, {Stevens}, {Takalo}, {Tavecchio},
  {Ter{\"a}sranta}, {Thompson}, {Tornikoski}, {Tosti}, {Ungerechts}, {Urry},
  {Valtaoja}, {Villata}, {Wagner}, {Wehrle}, \& {Wilson}}]{Hartman01}
{Hartman}, R.~C., {B{\"o}ttcher}, M., {Aldering}, G., {et~al.} 2001, \apj, 553,
  683

\bibitem[{{Healey} {et~al.}(2007){Healey}, {Romani}, {Taylor}, {Sadler},
  {Ricci}, {Murphy}, {Ulvestad}, \& {Winn}}]{Healey07}
{Healey}, S.~E., {Romani}, R.~W., {Taylor}, G.~B., {et~al.} 2007, \apjs, 171,
  61

\bibitem[{{Hervet} {et~al.}(2019){Hervet}, {Williams}, {Falcone}, \&
  {Kaur}}]{Hervet19}
{Hervet}, O., {Williams}, D.~A., {Falcone}, A.~D., \& {Kaur}, A. 2019, \apj,
  877, 26

\bibitem[{{Impey} \& {Neugebauer}(1988)}]{Impey88}
{Impey}, C.~D. \& {Neugebauer}, G. 1988, \aj, 95, 307

\bibitem[{{Isobe} {et~al.}(2010){Isobe}, {Sugimori}, {Kawai}, {Ueda}, {Negoro},
  {Sugizaki}, {Matsuoka}, {Daikyuji}, {Eguchi}, \& {Hiroi}}]{Isobe10}
{Isobe}, N., {Sugimori}, K., {Kawai}, N., {et~al.} 2010, \pasj, 62, L55

\bibitem[{{Jorstad} {et~al.}(2001){Jorstad}, {Marscher}, {Mattox}, {Wehrle},
  {Bloom}, \& {Yurchenko}}]{Jorstad01}
{Jorstad}, S.~G., {Marscher}, A.~P., {Mattox}, J.~R., {et~al.} 2001, \apjs,
  134, 181

\bibitem[{{Kalberla} {et~al.}(2005){Kalberla}, {Burton}, {Hartmann}, {Arnal},
  {Bajaja}, {Morras}, \& {P{\"o}ppel}}]{Kalberla05}
{Kalberla}, P.~M.~W., {Burton}, W.~B., {Hartmann}, D., {et~al.} 2005, \aap,
  440, 775

\bibitem[{{Kataoka} {et~al.}(2012){Kataoka}, {Yatsu}, {Kawai}, {Urata},
  {Cheung}, {Takahashi}, {Maeda}, {Totani}, {Makiya}, {Hanayama}, {Miyaji}, \&
  {Tsai}}]{Kataoka12}
{Kataoka}, J., {Yatsu}, Y., {Kawai}, N., {et~al.} 2012, \apj, 757, 176

\bibitem[{{Kaur} {et~al.}(2017){Kaur}, {Chandra}, {Baliyan}, {Sameer}, \&
  {Ganesh}}]{Kaur17}
{Kaur}, N., {Chandra}, S., {Baliyan}, K.~S., {Sameer}, \& {Ganesh}, S. 2017,
  \apj, 846, 158

\bibitem[{{Landi} {et~al.}(2015){Landi}, {Bassani}, {Stephen}, {Masetti},
  {Malizia}, \& {Ubertini}}]{Landi15}
{Landi}, R., {Bassani}, L., {Stephen}, J.~B., {et~al.} 2015, \aap, 581, A57

\bibitem[{{Landoni} {et~al.}(2015){Landoni}, {Massaro}, {Paggi}, {D'Abrusco},
  {Milisavljevic}, {Masetti}, {Smith}, {Tosti}, {Chomiuk}, {Strader}, \&
  {Cheung}}]{Opt3}
{Landoni}, M., {Massaro}, F., {Paggi}, A., {et~al.} 2015, \aj, 149, 163

\bibitem[{{Laurino} {et~al.}(2011){Laurino}, {D'Abrusco}, {Longo}, \&
  {Riccio}}]{Laurino11}
{Laurino}, O., {D'Abrusco}, R., {Longo}, G., \& {Riccio}, G. 2011, \mnras, 418,
  2165

\bibitem[{{Le{\'o}n-Tavares} {et~al.}(2011){Le{\'o}n-Tavares}, {Valtaoja},
  {Chavushyan}, {Tornikoski}, {A{\~n}orve}, {Nieppola}, \&
  {L{\"a}hteenm{\"a}ki}}]{LeonTavares11}
{Le{\'o}n-Tavares}, J., {Valtaoja}, E., {Chavushyan}, V.~H., {et~al.} 2011,
  \mnras, 411, 1127

\bibitem[{{Lico} {et~al.}(2014){Lico}, {Giroletti}, {Orienti}, {G{\'o}mez},
  {Casadio}, {D'Ammando}, {Blasi}, {Cotton}, {Edwards}, {Fuhrmann}, {Jorstad},
  {Kino}, {Kovalev}, {Krichbaum}, {Marscher}, {Paneque}, {Piner}, \&
  {Sokolovsky}}]{Lico14}
{Lico}, R., {Giroletti}, M., {Orienti}, M., {et~al.} 2014, \aap, 571, A54

\bibitem[{{Lister} {et~al.}(2013){Lister}, {Aller}, {Aller}, {Homan},
  {Kellermann}, {Kovalev}, {Pushkarev}, {Richards}, {Ros}, \&
  {Savolainen}}]{Lister13}
{Lister}, M.~L., {Aller}, M.~F., {Aller}, H.~D., {et~al.} 2013, \aj, 146, 120

\bibitem[{{Lister} {et~al.}(2019){Lister}, {Homan}, {Hovatta}, {Kellermann},
  {Kiehlmann}, {Kovalev}, {Max-Moerbeck}, {Pushkarev}, {Readhead}, \&
  {Ros}}]{Lister19}
{Lister}, M.~L., {Homan}, D.~C., {Hovatta}, T., {et~al.} 2019, \apj, 874, 43

\bibitem[{{Mahony} {et~al.}(2010){Mahony}, {Sadler}, {Murphy}, {Ekers},
  {Edwards}, \& {Massardi}}]{Mahony10}
{Mahony}, E.~K., {Sadler}, E.~M., {Murphy}, T., {et~al.} 2010, \apj, 718, 587

\bibitem[{{Mao} {et~al.}(2016){Mao}, {Urry}, {Massaro}, {Paggi}, {Cauteruccio},
  \& {K{\"u}nzel}}]{Mao16}
{Mao}, P., {Urry}, C.~M., {Massaro}, F., {et~al.} 2016, \apjs, 224, 26

\bibitem[{{Maraschi} {et~al.}(1992){Maraschi}, {Ghisellini}, \&
  {Celotti}}]{Maraschi92}
{Maraschi}, L., {Ghisellini}, G., \& {Celotti}, A. 1992, \apjl, 397, L5

\bibitem[{{Marchesini} {et~al.}(2016){Marchesini}, {Masetti}, {Chavushyan},
  {Cellone}, {Andruchow}, {Bassani}, {Bazzano}, {Jim{\'e}nez-Bail{\'o}n},
  {Landi}, {Malizia}, {Palazzi}, {Pati{\~n}o-{\'A}lvarez},
  {Rodr{\'{\i}}guez-Castillo}, {Stephen}, \& {Ubertini}}]{Marchesini16}
{Marchesini}, E.~J., {Masetti}, N., {Chavushyan}, V., {et~al.} 2016, \aap, 596,
  A10

\bibitem[{{Marchesini} {et~al.}(2019{\natexlab{a}}){Marchesini},
  {Pe{\~n}a-Herazo}, {{\'A}lvarez Crespo}, {Ricci}, {Negro}, {Milisavljevic},
  {Massaro}, {Masetti}, {Landoni}, {Chavushyan}, {D'Abrusco},
  {Jim{\'e}nez-Bail{\'o}n}, {La Franca}, {Paggi}, {Smith}, \&
  {Tosti}}]{Marchesini19}
{Marchesini}, E.~J., {Pe{\~n}a-Herazo}, H.~A., {{\'A}lvarez Crespo}, N.,
  {et~al.} 2019{\natexlab{a}}, \apss, 364, 5

\bibitem[{{Marchesini} {et~al.}(2019{\natexlab{b}}){Marchesini},
  {Pe{\~n}a-Herazo}, {{\'A}lvarez Crespo}, {Ricci}, {Negro}, {Milisavljevic},
  {Massaro}, {Masetti}, {Landoni}, {Chavushyan}, {D'Abrusco},
  {Jim{\'e}nez-Bail{\'o}n}, {La Franca}, {Paggi}, {Smith}, \& {Tosti}}]{Opt8}
{Marchesini}, E.~J., {Pe{\~n}a-Herazo}, H.~A., {{\'A}lvarez Crespo}, N.,
  {et~al.} 2019{\natexlab{b}}, \apss, 364, 5

\bibitem[{{Maselli} {et~al.}(2010){Maselli}, {Massaro}, {Nesci}, {Sclavi},
  {Rossi}, \& {Giommi}}]{Maselli10a}
{Maselli}, A., {Massaro}, E., {Nesci}, R., {et~al.} 2010, \aap, 512, A74

\bibitem[{{Maselli} {et~al.}(2013){Maselli}, {Massaro}, {Cusumano},
  {D'Abrusco}, {La Parola}, {Paggi}, {Segreto}, {Smith}, \&
  {Tosti}}]{Maselli13}
{Maselli}, A., {Massaro}, F., {Cusumano}, G., {et~al.} 2013, \apjs, 206, 17

\bibitem[{{Masetti} {et~al.}(2013){Masetti}, {Sbarufatti}, {Parisi},
  {Jim{\'e}nez-Bail{\'o}n}, {Chavushyan}, {Vogt}, {Sguera}, {Stephen},
  {Palazzi}, {Bassani}, {Bazzano}, {Fiocchi}, {Galaz}, {Landi}, {Malizia},
  {Minniti}, {Morelli}, \& {Ubertini}}]{Masetti13b}
{Masetti}, N., {Sbarufatti}, B., {Parisi}, P., {et~al.} 2013, \aap, 559, A58

\bibitem[{{Massaro} {et~al.}(2015{\natexlab{a}}){Massaro}, {Maselli}, {Leto},
  {Marchegiani}, {Perri}, {Giommi}, \& {Piranomonte}}]{Massaro15}
{Massaro}, E., {Maselli}, A., {Leto}, C., {et~al.} 2015{\natexlab{a}}, \apss,
  357, 75

\bibitem[{{Massaro} {et~al.}(2006){Massaro}, {Tramacere}, {Perri}, {Giommi}, \&
  {Tosti}}]{Massaro06}
{Massaro}, E., {Tramacere}, A., {Perri}, M., {Giommi}, P., \& {Tosti}, G. 2006,
  \aap, 448, 861

\bibitem[{{Massaro} {et~al.}(2016){Massaro}, {{\'A}lvarez Crespo}, {D'Abrusco},
  {Land oni}, {Masetti}, {Ricci}, {Milisavljevic}, {Paggi}, {Chavushyan}, \&
  {Jim{\'e}nez-Bail{\'o}n}}]{Massaro16b}
{Massaro}, F., {{\'A}lvarez Crespo}, N., {D'Abrusco}, R., {et~al.} 2016, \apss,
  361, 337

\bibitem[{{Massaro} \& {D'Abrusco}(2016)}]{Massaro16}
{Massaro}, F. \& {D'Abrusco}, R. 2016, \apj, 827, 67

\bibitem[{{Massaro} {et~al.}(2011{\natexlab{a}}){Massaro}, {D'Abrusco},
  {Ajello}, {Grindlay}, \& {Smith}}]{Massaro11}
{Massaro}, F., {D'Abrusco}, R., {Ajello}, M., {Grindlay}, J.~E., \& {Smith},
  H.~A. 2011{\natexlab{a}}, \apjl, 740, L48

\bibitem[{{Massaro} {et~al.}(2013{\natexlab{a}}){Massaro}, {D'Abrusco},
  {Giroletti}, {Paggi}, {Masetti}, {Tosti}, {Nori}, \& {Funk}}]{Massaro13b}
{Massaro}, F., {D'Abrusco}, R., {Giroletti}, M., {et~al.} 2013{\natexlab{a}},
  \apjs, 207, 4

\bibitem[{{Massaro} {et~al.}(2013{\natexlab{b}}){Massaro}, {D'Abrusco},
  {Giroletti}, {Paggi}, {Masetti}, {Tosti}, {Nori}, \& {Funk}}]{Coso2}
{Massaro}, F., {D'Abrusco}, R., {Giroletti}, M., {et~al.} 2013{\natexlab{b}},
  \apjs, 207, 4

\bibitem[{{Massaro} {et~al.}(2015{\natexlab{b}}){Massaro}, {D'Abrusco},
  {Landoni}, {Paggi}, {Masetti}, {Giroletti}, {Ot{\'\i}-Floranes},
  {Chavushyan}, {Jim{\'e}nez-Bail{\'o}n}, \&
  {Pati{\~n}o-{\'A}lvarez}}]{Massaro15c}
{Massaro}, F., {D'Abrusco}, R., {Landoni}, M., {et~al.} 2015{\natexlab{b}},
  \apjs, 217, 2

\bibitem[{{Massaro} {et~al.}(2013{\natexlab{c}}){Massaro}, {D'Abrusco},
  {Paggi}, {Masetti}, {Giroletti}, {Tosti}, {Smith}, \& {Funk}}]{Massaro13d}
{Massaro}, F., {D'Abrusco}, R., {Paggi}, A., {et~al.} 2013{\natexlab{c}},
  \apjs, 209, 10

\bibitem[{{Massaro} {et~al.}(2012{\natexlab{a}}){Massaro}, {D'Abrusco},
  {Tosti}, {Ajello}, {Gasparrini}, {Grindlay}, \& {Smith}}]{Massaro12b}
{Massaro}, F., {D'Abrusco}, R., {Tosti}, G., {et~al.} 2012{\natexlab{a}}, \apj,
  750, 138

\bibitem[{{Massaro} {et~al.}(2012{\natexlab{b}}){Massaro}, {D'Abrusco},
  {Tosti}, {Ajello}, {Paggi}, \& {Gasparrini}}]{Massaro12c}
{Massaro}, F., {D'Abrusco}, R., {Tosti}, G., {et~al.} 2012{\natexlab{b}}, \apj,
  752, 61

\bibitem[{{Massaro} {et~al.}(2008{\natexlab{a}}){Massaro}, {Giommi}, {Tosti},
  {Cassetti}, {Nesci}, {Perri}, {Burrows}, \& {Gerehls}}]{Massaro08b}
{Massaro}, F., {Giommi}, P., {Tosti}, G., {et~al.} 2008{\natexlab{a}}, \aap,
  489, 1047

\bibitem[{{Massaro} {et~al.}(2013{\natexlab{d}}){Massaro}, {Giroletti},
  {Paggi}, {D'Abrusco}, {Tosti}, \& {Funk}}]{Massaro13c}
{Massaro}, F., {Giroletti}, M., {Paggi}, A., {et~al.} 2013{\natexlab{d}},
  \apjs, 208, 15

\bibitem[{{Massaro} {et~al.}(2011{\natexlab{b}}){Massaro}, {Harris}, \&
  {Cheung}}]{Massaro11c}
{Massaro}, F., {Harris}, D.~E., \& {Cheung}, C.~C. 2011{\natexlab{b}}, \apjs,
  197, 24

\bibitem[{{Massaro} {et~al.}(2015{\natexlab{c}}){Massaro}, {Landoni},
  {D'Abrusco}, {Milisavljevic}, {Paggi}, {Masetti}, {Smith}, \& {Tosti}}]{Opt2}
{Massaro}, F., {Landoni}, M., {D'Abrusco}, R., {et~al.} 2015{\natexlab{c}},
  \aap, 575, A124

\bibitem[{{Massaro} {et~al.}(2014){Massaro}, {Masetti}, {D'Abrusco}, {Paggi},
  \& {Funk}}]{Coso1}
{Massaro}, F., {Masetti}, N., {D'Abrusco}, R., {Paggi}, A., \& {Funk}, S. 2014,
  \aj, 148, 66

\bibitem[{{Massaro} {et~al.}(2011{\natexlab{c}}){Massaro}, {Paggi}, \&
  {Cavaliere}}]{Massaro11a}
{Massaro}, F., {Paggi}, A., \& {Cavaliere}, A. 2011{\natexlab{c}}, \apjl, 742,
  L32

\bibitem[{{Massaro} {et~al.}(2011{\natexlab{d}}){Massaro}, {Paggi}, {Elvis}, \&
  {Cavaliere}}]{Massaro11b}
{Massaro}, F., {Paggi}, A., {Elvis}, M., \& {Cavaliere}, A. 2011{\natexlab{d}},
  \apj, 739, 73

\bibitem[{{Massaro} {et~al.}(2013{\natexlab{e}}){Massaro}, {Paggi}, {Errando},
  {D'Abrusco}, {Masetti}, {Tosti}, \& {Funk}}]{Massaro13}
{Massaro}, F., {Paggi}, A., {Errando}, M., {et~al.} 2013{\natexlab{e}}, \apjs,
  207, 16

\bibitem[{{Massaro} {et~al.}(2015{\natexlab{d}}){Massaro}, {Thompson}, \&
  {Ferrara}}]{Massaro15b}
{Massaro}, F., {Thompson}, D.~J., \& {Ferrara}, E.~C. 2015{\natexlab{d}},
  \aapr, 24, 2

\bibitem[{{Massaro} {et~al.}(2008{\natexlab{b}}){Massaro}, {Tramacere},
  {Cavaliere}, {Perri}, \& {Giommi}}]{Massaro08}
{Massaro}, F., {Tramacere}, A., {Cavaliere}, A., {Perri}, M., \& {Giommi}, P.
  2008{\natexlab{b}}, \aap, 478, 395

\bibitem[{{Massaro} {et~al.}(2012{\natexlab{c}}){Massaro}, {Tremblay},
  {Harris}, {Kharb}, {Axon}, {Balmaverde}, {Baum}, {Capetti}, {Chiaberge}, \&
  {Gilli}}]{Massaro12d}
{Massaro}, F., {Tremblay}, G.~R., {Harris}, D.~E., {et~al.} 2012{\natexlab{c}},
  \apjs, 203, 31

\bibitem[{{Mastichiadis} \& {Kirk}(1997)}]{Mastichiadis97}
{Mastichiadis}, A. \& {Kirk}, J.~G. 1997, \aap, 320, 19

\bibitem[{{Mattox} \& {Ormes}(2002)}]{Mattox02}
{Mattox}, J.~R. \& {Ormes}, J.~F. 2002, in American Astronomical Society
  Meeting Abstracts, Vol. 201, 147.02

\bibitem[{{Mauch} {et~al.}(2003){Mauch}, {Murphy}, {Buttery}, {Curran},
  {Hunstead}, {Piestrzynski}, {Robertson}, \& {Sadler}}]{Mauch03}
{Mauch}, T., {Murphy}, T., {Buttery}, H.~J., {et~al.} 2003, \mnras, 342, 1117

\bibitem[{{Mirabal}(2009)}]{Mirabal09}
{Mirabal}, N. 2009, arXiv e-prints, arXiv:0908.1389

\bibitem[{{Moretti} {et~al.}(2004){Moretti}, {Campana}, {Tagliaferri}, {Abbey},
  {Ambrosi}, {Angelini}, {Beardmore}, {Br{\"a}uninger}, {Burkert}, {Burrows},
  {Capalbi}, {Chincarini}, {Citterio}, {Cusumano}, {Freyberg}, {Giommi},
  {Hartner}, {Hill}, {Mori}, {Morris}, {Mukerjee}, {Nousek}, {Osborne},
  {Short}, {Tamburelli}, {Watson}, \& {Wells}}]{Moretti04}
{Moretti}, A., {Campana}, S., {Tagliaferri}, G., {et~al.} 2004, in \procspie,
  Vol. 5165, X-Ray and Gamma-Ray Instrumentation for Astronomy XIII, ed. K.~A.
  {Flanagan} \& O.~H.~W. {Siegmund}, 232--240

\bibitem[{{Mukai}(1993)}]{Mukai93}
{Mukai}, K. 1993, Legacy, vol.~3, p.21-31, 3, 21

\bibitem[{{Nori} {et~al.}(2014){Nori}, {Giroletti}, {Massaro}, {D'Abrusco},
  {Paggi}, {Tosti}, \& {Funk}}]{Nori14}
{Nori}, M., {Giroletti}, M., {Massaro}, F., {et~al.} 2014, \apjs, 212, 3

\bibitem[{{Paggi} {et~al.}(2011){Paggi}, {Cavaliere}, {Vittorini}, {D'Ammando},
  \& {Tavani}}]{Paggi11}
{Paggi}, A., {Cavaliere}, A., {Vittorini}, V., {D'Ammando}, F., \& {Tavani}, M.
  2011, \apj, 736, 128

\bibitem[{{Paggi} {et~al.}(2009{\natexlab{a}}){Paggi}, {Cavaliere},
  {Vittorini}, \& {Tavani}}]{Paggi09b}
{Paggi}, A., {Cavaliere}, A., {Vittorini}, V., \& {Tavani}, M.
  2009{\natexlab{a}}, \aap, 508, L31

\bibitem[{{Paggi} {et~al.}(2013){Paggi}, {Massaro}, {D'Abrusco}, {Smith},
  {Masetti}, {Giroletti}, {Tosti}, \& {Funk}}]{Paggi13}
{Paggi}, A., {Massaro}, F., {D'Abrusco}, R., {et~al.} 2013, \apjs, 209, 9

\bibitem[{{Paggi} {et~al.}(2009{\natexlab{b}}){Paggi}, {Massaro}, {Vittorini},
  {Cavaliere}, {D'Ammando}, {Vagnetti}, \& {Tavani}}]{Paggi09a}
{Paggi}, A., {Massaro}, F., {Vittorini}, V., {et~al.} 2009{\natexlab{b}}, \aap,
  504, 821

\bibitem[{{Paggi} {et~al.}(2014){Paggi}, {Milisavljevic}, {Masetti},
  {Jim{\'e}nez-Bail{\'o}n}, {Chavushyan}, {D'Abrusco}, {Massaro}, {Giroletti},
  {Smith}, {Margutti}, {Tosti}, {Mart{\'{\i}}nez-Galarza},
  {Ot{\'{\i}}-Floranes}, {Landoni}, {Grindlay}, \& {Funk}}]{Opt1}
{Paggi}, A., {Milisavljevic}, D., {Masetti}, N., {et~al.} 2014, \aj, 147, 112

\bibitem[{{Paiano} {et~al.}(2017){Paiano}, {Franceschini}, \&
  {Stamerra}}]{Paiano17a}
{Paiano}, S., {Franceschini}, A., \& {Stamerra}, A. 2017, \mnras, 468, 4902

\bibitem[{{Pandey} {et~al.}(2017){Pandey}, {Gupta}, \& {Wiita}}]{Pandey17}
{Pandey}, A., {Gupta}, A.~C., \& {Wiita}, P.~J. 2017, \apj, 841, 123

\bibitem[{{Pe{\~n}a-Herazo} {et~al.}(2017{\natexlab{a}}){Pe{\~n}a-Herazo},
  {Marchesini}, {{\'A}lvarez Crespo}, {Ricci}, {Massaro}, {Chavushyan},
  {Landoni}, {Strader}, {Chomiuk}, {Cheung}, {Masetti},
  {Jim{\'e}nez-Bail{\'o}n}, {D'Abrusco}, {Paggi}, {Milisavljevic}, {La Franca},
  {Smith}, \& {Tosti}}]{PenaHerazo17}
{Pe{\~n}a-Herazo}, H.~A., {Marchesini}, E.~J., {{\'A}lvarez Crespo}, N.,
  {et~al.} 2017{\natexlab{a}}, \apss, 362, 228

\bibitem[{{Pe{\~n}a-Herazo} {et~al.}(2017{\natexlab{b}}){Pe{\~n}a-Herazo},
  {Marchesini}, {{\'A}lvarez Crespo}, {Ricci}, {Massaro}, {Chavushyan},
  {Landoni}, {Strader}, {Chomiuk}, {Cheung}, {Masetti},
  {Jim{\'e}nez-Bail{\'o}n}, {D'Abrusco}, {Paggi}, {Milisavljevic}, {La Franca},
  {Smith}, \& {Tosti}}]{Opt7}
{Pe{\~n}a-Herazo}, H.~A., {Marchesini}, E.~J., {{\'A}lvarez Crespo}, N.,
  {et~al.} 2017{\natexlab{b}}, \apss, 362, 228

\bibitem[{{Pe{\~n}a-Herazo} {et~al.}(2019){Pe{\~n}a-Herazo}, {Massaro},
  {Chavushyan}, {Marchesini}, {Paggi}, {Landoni}, {Masetti}, {Ricci},
  {D'Abrusco}, \& {Milisavljevic}}]{Opt9}
{Pe{\~n}a-Herazo}, H.~A., {Massaro}, F., {Chavushyan}, V., {et~al.} 2019,
  \apss, 364, 85

\bibitem[{{Pian} {et~al.}(1998){Pian}, {Vacanti}, {Tagliaferri}, {Ghisellini},
  {Maraschi}, {Treves}, {Urry}, {Fiore}, {Giommi}, {Palazzi}, {Chiappetti}, \&
  {Sambruna}}]{Pian98}
{Pian}, E., {Vacanti}, G., {Tagliaferri}, G., {et~al.} 1998, \apjl, 492, L17

\bibitem[{{Piner} \& {Edwards}(2014)}]{Piner14}
{Piner}, B.~G. \& {Edwards}, P.~G. 2014, \apj, 797, 25

\bibitem[{{Piner} \& {Edwards}(2018)}]{Piner18}
{Piner}, B.~G. \& {Edwards}, P.~G. 2018, \apj, 853, 68

\bibitem[{{Plotkin} {et~al.}(2011){Plotkin}, {Markoff}, {Trager}, \&
  {Anderson}}]{Plotkin11}
{Plotkin}, R.~M., {Markoff}, S., {Trager}, S.~C., \& {Anderson}, S.~F. 2011,
  \mnras, 413, 805

\bibitem[{{Ricci} {et~al.}(2015){Ricci}, {Massaro}, {Landoni}, {D'Abrusco},
  {Milisavljevic}, {Stern}, {Masetti}, {Paggi}, {Smith}, \& {Tosti}}]{Opt4}
{Ricci}, F., {Massaro}, F., {Landoni}, M., {et~al.} 2015, \aj, 149, 160

\bibitem[{{Romero} {et~al.}(2002){Romero}, {Cellone}, {Combi}, \&
  {Andruchow}}]{Romero02}
{Romero}, G.~E., {Cellone}, S.~A., {Combi}, J.~A., \& {Andruchow}, I. 2002,
  \aap, 390, 431

\bibitem[{{Sambruna} {et~al.}(1996){Sambruna}, {Maraschi}, \&
  {Urry}}]{Sambruna96}
{Sambruna}, R.~M., {Maraschi}, L., \& {Urry}, C.~M. 1996, \apj, 463, 444

\bibitem[{{Sbarrato} {et~al.}(2012){Sbarrato}, {Ghisellini}, {Maraschi}, \&
  {Colpi}}]{Sbarrato12}
{Sbarrato}, T., {Ghisellini}, G., {Maraschi}, L., \& {Colpi}, M. 2012, \mnras,
  421, 1764

\bibitem[{{Shaw} {et~al.}(2012){Shaw}, {Romani}, {Cotter}, {Healey},
  {Michelson}, {Readhead}, {Richards}, {Max-Moerbeck}, {King}, \&
  {Potter}}]{Shaw12}
{Shaw}, M.~S., {Romani}, R.~W., {Cotter}, G., {et~al.} 2012, \apj, 748, 49

\bibitem[{{Sikora} {et~al.}(2013){Sikora}, {Janiak}, {Nalewajko}, {Madejski},
  \& {Moderski}}]{Sikora13}
{Sikora}, M., {Janiak}, M., {Nalewajko}, K., {Madejski}, G.~M., \& {Moderski},
  R. 2013, \apj, 779, 68

\bibitem[{{Singh} \& {Garmire}(1985)}]{Singh85}
{Singh}, K.~P. \& {Garmire}, G.~P. 1985, \apj, 297, 199

\bibitem[{{Stecker} {et~al.}(1993){Stecker}, {Salamon}, \&
  {Malkan}}]{Stecker93}
{Stecker}, F.~W., {Salamon}, M.~H., \& {Malkan}, M.~A. 1993, \apj, 410, L71

\bibitem[{{Stevens} {et~al.}(1994){Stevens}, {Litchfield}, {Robson}, {Hughes},
  {Gear}, {Terasranta}, {Valtaoja}, \& {Tornikoski}}]{Stevens94}
{Stevens}, J.~A., {Litchfield}, S.~J., {Robson}, E.~I., {et~al.} 1994, \apj,
  437, 91

\bibitem[{{Stickel} {et~al.}(1991){Stickel}, {Padovani}, {Urry}, {Fried}, \&
  {Kuehr}}]{Stickel91}
{Stickel}, M., {Padovani}, P., {Urry}, C.~M., {Fried}, J.~W., \& {Kuehr}, H.
  1991, \apj, 374, 431

\bibitem[{{Stroh} \& {Falcone}(2013)}]{Stroh13}
{Stroh}, M.~C. \& {Falcone}, A.~D. 2013, \apjs, 207, 28

\bibitem[{{Tavecchio} \& {Ghisellini}(2016)}]{Tavecchio16}
{Tavecchio}, F. \& {Ghisellini}, G. 2016, \mnras, 456, 2374

\bibitem[{{Tavecchio} {et~al.}(2011){Tavecchio}, {Ghisellini}, {Bonnoli}, \&
  {Foschini}}]{Tavecchio11}
{Tavecchio}, F., {Ghisellini}, G., {Bonnoli}, G., \& {Foschini}, L. 2011,
  \mnras, 414, 3566

\bibitem[{{Taylor} {et~al.}(2007){Taylor}, {Healey}, {Helmboldt}, {Tremblay},
  {Fassnacht}, {Walker}, {Sjouwerman}, {Pearson}, {Readhead}, \&
  {Weintraub}}]{Taylor07}
{Taylor}, G.~B., {Healey}, S.~E., {Helmboldt}, J.~F., {et~al.} 2007, \apj, 671,
  1355

\bibitem[{{Taylor}(2005)}]{Taylor05}
{Taylor}, M.~B. 2005, in Astronomical Society of the Pacific Conference Series,
  Vol. 347, Astronomical Data Analysis Software and Systems XIV, ed.
  P.~{Shopbell}, M.~{Britton}, \& R.~{Ebert}, 29

\bibitem[{{Taylor}(2006)}]{Taylor06}
{Taylor}, M.~B. 2006, in Astronomical Society of the Pacific Conference Series,
  Vol. 351, Astronomical Data Analysis Software and Systems XV, ed.
  C.~{Gabriel}, C.~{Arviset}, D.~{Ponz}, \& S.~{Enrique}, 666

\bibitem[{{Tchekhovskoy} {et~al.}(2009){Tchekhovskoy}, {McKinney}, \&
  {Narayan}}]{Tchekhovskoy09}
{Tchekhovskoy}, A., {McKinney}, J.~C., \& {Narayan}, R. 2009, \apj, 699, 1789

\bibitem[{{Tramacere} {et~al.}(2011){Tramacere}, {Massaro}, \&
  {Taylor}}]{Tramacere11}
{Tramacere}, A., {Massaro}, E., \& {Taylor}, A.~M. 2011, \apj, 739, 66

\bibitem[{{Tramacere} {et~al.}(2007){Tramacere}, {Massaro}, \&
  {Cavaliere}}]{Tramacere07}
{Tramacere}, A., {Massaro}, F., \& {Cavaliere}, A. 2007, \aap, 466, 521

\bibitem[{{Voges} {et~al.}(1999){Voges}, {Aschenbach}, {Boller},
  {Br{\"a}uninger}, {Briel}, {Burkert}, {Dennerl}, {Englhauser}, {Gruber},
  {Haberl}, {Hartner}, {Hasinger}, {K{\"u}rster}, {Pfeffermann}, {Pietsch},
  {Predehl}, {Rosso}, {Schmitt}, {Tr{\"u}mper}, \& {Zimmermann}}]{Voges99}
{Voges}, W., {Aschenbach}, B., {Boller}, T., {et~al.} 1999, \aap, 349, 389

\bibitem[{{Wehrle} {et~al.}(1998){Wehrle}, {Pian}, {Urry}, {Maraschi},
  {McHardy}, {Lawson}, {Ghisellini}, {Hartman}, {Madejski}, {Makino},
  {Marscher}, {Wagner}, {Webb}, {Aldering}, {Aller}, {Aller}, {Backman},
  {Balonek}, {Boltwood}, {Bonnell}, {Caplinger}, {Celotti}, {Collmar},
  {Dalton}, {Drucker}, {Falomo}, {Fichtel}, {Freudling}, {Gear}, {Gonzales},
  {Hall}, {Inoue}, {Johnson}, {Kazanas}, {Kidger}, {Kii}, {Kollgaard}, {Kondo},
  {Kurfess}, {Lin}, {McCollum}, {McNaron-Brown}, {Nagase}, {Nair}, {Penton},
  {Pesce}, {Pohl}, {Raiteri}, {Renda}, {Robson}, {Sambruna}, {Schirmer},
  {Shrader}, {Sikora}, {Sillanp{\"a}{\"a}}, {Smith}, {Stevens}, {Stocke},
  {Takalo}, {Ter{\"a}sranta}, {Thompson}, {Thompson}, {Tornikoski}, {Tosti},
  {Treves}, {Turcotte}, {Unwin}, {Valtaoja}, {Villata}, {Xu}, {Yamashita}, \&
  {Zook}}]{Wehrle98}
{Wehrle}, A.~E., {Pian}, E., {Urry}, C.~M., {et~al.} 1998, \apj, 497, 178

\bibitem[{{White} {et~al.}(1997){White}, {Becker}, {Helfand}, \&
  {Gregg}}]{White97}
{White}, R.~L., {Becker}, R.~H., {Helfand}, D.~J., \& {Gregg}, M.~D. 1997,
  \apj, 475, 479

\bibitem[{{Woo} {et~al.}(2005){Woo}, {Urry}, {van der Marel}, {Lira}, \&
  {Maza}}]{Woo05}
{Woo}, J.-H., {Urry}, C.~M., {van der Marel}, R.~P., {Lira}, P., \& {Maza}, J.
  2005, \apj, 631, 762

\bibitem[{{Wright} {et~al.}(2010){Wright}, {Eisenhardt}, {Mainzer}, {Ressler},
  {Cutri}, {Jarrett}, {Kirkpatrick}, {Padgett}, {McMillan}, {Skrutskie},
  {Stanford}, {Cohen}, {Walker}, {Mather}, {Leisawitz}, {Gautier}, {McLean},
  {Benford}, {Lonsdale}, {Blain}, {Mendez}, {Irace}, {Duval}, {Liu}, {Royer},
  {Heinrichsen}, {Howard}, {Shannon}, {Kendall}, {Walsh}, {Larsen}, {Cardon},
  {Schick}, {Schwalm}, {Abid}, {Fabinsky}, {Naes}, \& {Tsai}}]{Wright10}
{Wright}, E.~L., {Eisenhardt}, P.~R.~M., {Mainzer}, A.~K., {et~al.} 2010, \aj,
  140, 1868

\bibitem[{{Xiong} \& {Zhang}(2014)}]{Xiong14}
{Xiong}, D.~R. \& {Zhang}, X. 2014, \mnras, 441, 3375

\end{thebibliography}
\bibliographystyle{aa.bst}
\newpage

\end{document}